\let\latexaddtocontents\addtocontents
\documentclass[a4paper,twocolumn,10pt, unpublished]{quantumarticle}
\let\addtocontents\latexaddtocontents

\pdfoutput=1

\usepackage[utf8]{inputenc}
\usepackage{booktabs}
\usepackage{amsmath}
\usepackage{fullpage}
\usepackage{amsthm}
\usepackage{amsfonts}
\usepackage{amssymb}
\usepackage{mathtools}
\usepackage{braket}
\usepackage{algorithm2e}
\usepackage{multirow}
\usepackage{algpseudocode}
\usepackage{appendix}
\usepackage{xcolor}
\usepackage[official]{eurosym}
\usepackage{fontawesome}
\usepackage{pgfplots}
\pgfplotsset{compat=1.3}
\usepackage{tikz}
\usepackage{tabularx}
\usepackage{cellspace}
\usetikzlibrary{quantikz2}
\usetikzlibrary{tikzmark}
\usetikzlibrary{decorations.pathreplacing}
\usepackage{subcaption}
\usepackage{rotating}
\usepackage{hyperref}  
\usepackage[capitalize]{cleveref}  
\usepackage[style=numeric-comp,
backend=bibtex8,
doi=true,
isbn=false,
url=false,
eprint=false,
maxbibnames=5,
sorting=none]{biblatex}
\renewbibmacro{in:}{}
\crefdefaultlabelformat{#2#1#3}  
\crefformat{equation}{Eq.~#2#1#3}  

\addparagraphcolumntypes{X}

\definecolor{ForestGreen}{RGB}{34,139,34}
\definecolor{DarkRed}{RGB}{204,0,0}
\newcommand{\keepleave}{\iffalse}

\newcommand{\init}[0]{{\rm in}}
\newcommand{\coll}[0]{{\rm col}}
\newcommand{\need}[0]{{\rm need}}
\newcommand{\debtor}[0]{{\rm d}}
\newcommand{\creditor}[0]{{\rm c}}

\newcommand{\limit}[0]{{\rm lim}}
\newcommand{\trsh}[0]{{\rm min}}

\newcommand{\X}[0]{\mathbf{x}}
\newcommand{\x}[0]{x}
\newcommand{\xt}[0]{\x_{\transa}}
\newcommand{\Y}[0]{\mathbf{y}}
\newcommand{\yp}[0]{y_{\spl}}

\newcommand{\T}[0]{\mathbb{T}}
\newcommand{\transa}[0]{t}

\newcommand{\am}[0]{a}
\newcommand{\aIn}[0]{\am^\init}
\newcommand{\aCo}[0]{\am^\coll}
\newcommand{\aNe}[0]{\am^\need}
\newcommand{\aLi}[0]{\am^\limit}

\newcommand{\at}[0]{\am_{\transa}}

\newcommand{\q}[0]{q}
\newcommand{\qt}[0]{\q_{\transa}}
\newcommand{\qMi}[0]{\q^\trsh}
\newcommand{\qLi}[0]{\q^\limit}
\newcommand{\qIn}[0]{\q^\init}
\newcommand{\qCo}[0]{\q^\coll}

\newcommand{\ofun}[0]{\mathcal{F}}

\newcommand{\w}[0]{w}

\newcommand{\cbal}[0]{b}

\newcommand{\spos}[0]{p}

\newcommand{\secu}[0]{s}

\newcommand{\CMB}[0]{\mathbb{CMB}}
\newcommand{\cmb}[0]{m}
\newcommand{\SPL}[0]{\mathbb{SPL}}
\newcommand{\spl}[0]{l}
\newcommand{\lot}[0]{n}
\newcommand{\val}[0]{v}

\renewcommand{\tilde}{\widetilde}  
\renewcommand{\epsilon}{\varepsilon}  

\theoremstyle{plain}

\usepackage{dashrule}

\begin{document}

\title{Securities Transaction Settlement Optimization on superconducting quantum devices}
\author{Francesco Martini}
    \email{f\_martini+qc@outlook.it}
    \affiliation{Department of Computer Science, University of Verona, Verona, Italy}
\author{Daniele Lizzio Bosco}
    \affiliation{Department of Mathematics, Computer Science and Physics, University of Udine, Udine, Italy} 
\author{Carlo Barbanera}
    \affiliation{Directorate General for Markets and Payment Systems, Bank of Italy, Rome, Italy}
\author{Serena Bernardini}
    \affiliation{Directorate General for Markets and Payment Systems, Bank of Italy, Rome, Italy}
\author{Giacomo Ranieri}
    \affiliation{Data, AI and Technology, Intesa Sanpaolo, Torino, Italy}
\author{Francesca Cibrario}
    \affiliation{Data, AI and Technology, Intesa Sanpaolo, Torino, Italy}
\author{Davide Corbelletto}
    \affiliation{Data, AI and Technology, Intesa Sanpaolo, Torino, Italy}
\author{Giuseppe Bruno}
    \affiliation{Directorate General Economics, Statistics and Research, Bank of Italy, Rome, Italy}
\author{Alessandra Di Pierro}
  \affiliation{Department of Computer Science, University of Verona, Verona, Italy}
\author{Luca Dellantonio}
    \email{l.dellantonio@exeter.ac.uk}
    \affiliation{Department of Physics and Astronomy, University of Exeter, 
    Exeter EX4 4QL, United Kingdom}
\date{
\parbox{\linewidth}{
\centering
\endgraf \bigskip
\textbf{The views expressed here are the authors’ only and do not imply those of the Bank of Italy}. 
}
}

\begin{abstract}
We describe a quantum variational algorithm for securities transactions settlement optimization, based on a novel mathematical formalization of the problem that includes the most relevant constraints considered in the pan-European securities settlement platform \emph{TARGET2-Securities}. The proposed algorithm is designed for Noisy Intermediate-Scale Quantum devices, specifically targeting IBM’s superconducting qubit machines. We adopt non-linear activation functions to encode inequality constraints in the objective function of the problem, and design customized noise mitigation techniques to alleviate the effect of readout errors. We consider batches of up to 40 trades obtained from real transactional data to benchmark our algorithm on quantum hardware against classical and quantum-inspired solvers.
\end{abstract}

\maketitle

\section{Introduction}

Since Feynman's article in 1982 \cite{feynman1982simulating}, quantum computing has gained significant interest from academia and industry due to its potential to solve some classically hard problems \cite{shor1999polynomial, grover1996fast}. While quantum supremacy is still debated \cite{begusic2023fast,pednault2019leveraging,kalai2023questions,Liu2021closing}, the latest technological advances have enabled quantum computers to solve problems of practical relevance \cite{kim2023evidence, mcmahon2024improving}, including cryptography \cite{portmann2022security}, chemistry \cite{motta2022emerging,chan2024measurement,braine2021quantum,huber2024exponential} and finance \cite{chakrabarti2021threshold}. 

As hardware progresses, it is crucial to provide novel or refined applications that can run on the current generation of noisy intermediate-scale quantum (NISQ) devices \cite{preskill2018quantum}. 
In this work we analyze the pan European platform for security transaction settlement and develop a quantum algorithm for the \emph{Night-time Transactions Settlement Problem} (NTSP) \cite{alekseeva2020securities,T2SUDFS}, and show its behavior on the superconducting devices developed by IBM \cite{arute2019quantum}. Our demonstrations involve small-scale NTSP instances of up to $40$ transactions (mapped one-to-one with the qubits of the quantum computer). 

\begin{figure*}[ht!]
    \centering
    \includegraphics[width=\textwidth]{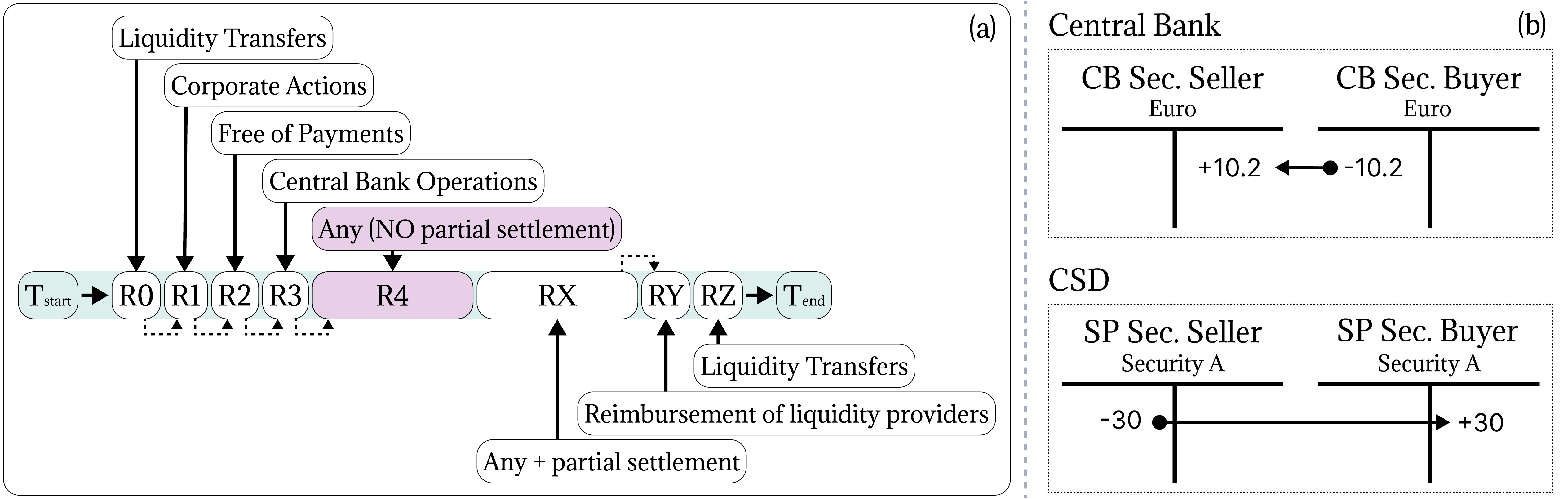}
    \caption{
    (a) Schematic diagram representing the sequence of optimization runs performed by T2S during the night-time cycles. Each run processes unsettled transactions from previous runs (dashed arrows), as well as transactions ``injected'' in T2S during that specific sequence run. We highlight in violet the run $4$, focus of the problem modelization provided in Sec.~\ref{sec:model}, during which no restrictions are placed on the types of transactions to be processed. Run $4$ must be completed within a $45$-minute timeframe \cite{alekseeva2020securities,T2SUDFS}. (b) T-accounts graphical representation of a DVP transaction involving the exchange of $30$ security of type A in exchange for $10.2$\euro{}. The sold securities are deducted from the security-seller's position and added to the security-buyer's position. Conversely, the $10.2$\euro{} from the security-buyer’s cash balance are transferred to the security-seller's DCA.
    }
    \label{fig:cycles_dvp}
\end{figure*}

Transaction settlement refers to a trade completion in which the buyer receives the purchased securities and the seller receives the agreed-upon payment. 
\emph{TARGET2-Securities} (T2S) \cite{T2SUDFS} is employed by Eurosystem, the monetary authority of the eurozone, to automate and optimize this task. Given a batch of transactions to process, during the Night-time Settlement phase (see Fig.~\ref{fig:cycles_dvp}) T2S improves the settlement of securities by maximizing the volume and amounts of eligible transactions while minimizing liquidity needs. 

The NTSP consists of finding the optimal subset of transactions (from a given batch) that the system can successfully settle.
It is modeled as a Mixed Integer Programming (MIP) problem \cite{fischetti2019introduction, alekseeva2020securities}, which is NP-complete 
\cite{karp2010reducibility, guntzer1998efficient}.
Due to the high volume of transactions processed and the computational complexity of the optimization tasks involved, T2S relies on heuristics to find viable solutions within the dedicated night-time settlement window (see Sec.~\ref{sec:STSO}). These heuristics typically yield suboptimal results, resulting in economic losses that arise when, e.g., transactions that could be settled are delayed or canceled. This highlights the need for more robust tools that could achieve better solutions for extremely large NTSP instances. 

In this work, we solve small-scale instances of the NTSP with a novel approach based on Variational Quantum Algorithms (VQAs) \cite{cerezo2021variational}. VQAs rely on a quantum-classical loop to determine the parameters of a circuit ansatz (i.e., parametrized quantum circuit) such that the quantum state produced encodes the desired solution.
The success of VQAs lies in their noise resilience \cite{Sharma2020noise}.
As such, despite the challenges in (hyper) parameter tuning and achieving reliable performance guarantees \cite{scriva2024challenges, de2023limitations, holmes2022connecting}, VQAs are valuable to understand the problems that current quantum devices can address \cite{haase2021resource,paulson2021simulating,ferguson2021measurement,chan2024measurement,gunderman2024minimal}. We establish the applicability of VQA to the NTSP by recasting this model as an unconstrained optimization task \cite{braine2021quantum, huber2024exponential}. We then choose a hardware-efficient ansatz structure that is applicable to larger and more complex problem instances.

The main contributions of this work are: 
\begin{enumerate}
    \item casting the NTSP \cite{alekseeva2020securities,T2SUDFS} problem in a form that captures the essential requirements addressed by T2S and is suitable for quantum algorithms; 
    \item designing a new VQA approach for objective functions encoding MIP problems by means of non-linear activation functions; 
    \item introducing a readout error mitigation that enables the convergence of the VQA for up to $20$ qubits.
\end{enumerate}
We evaluate the proposed variational architecture on gate-based superconducting quantum hardware, assessing its performance, limitations, and potential to compete with modern classical solvers.

The manuscript is organized as follows. In Sec.~\ref{sec:STSO}, we introduce the MIP formalization of the NTSP problem.
In Sec.~\ref{sec:methodology}, we detail each step of the proposed VQA, with a focus on the reformulation of the NTSP as an unconstrained problem and on the implemented custom noise mitigation technique. Experiments conducted on real quantum hardware evaluating the VQA are discussed in Sec.~\ref{sec:experiment}, while in Sec.~\ref{sec:conclusion} we present the conclusions and suggest future research directions.

\section{T2S Transaction Settlement Optimization Problem}
\label{sec:STSO}

NTSP solvers must ensure accurate, timely, and risk-free transfers of securities and funds between parties involved in the trades. Optimizing the securities transactions settlement problem is a complex task, demanding real-time processing of large volumes of data with strict execution constraints. Designed to meet these requirements, T2S \cite{T2SPrinciples} manages high transaction volumes involving central banks and Central Securities Depositories (CSDs: financial organizations that hold securities and facilitate digital ownership transfers) across both euro and non-euro areas. In this section, we first provide an overview of the NTSP problem, followed by the formalization of the model employed by the experiments in Sec.~\ref{sec:experiment}.

\subsection {Overview}
\label{sec:NTSP_overview}

T2S operates according to a schedule that can be divided into Night-time and Real-time Settlement periods. 
As shown in Fig.~\ref{fig:cycles_dvp}(a), we focus on the problem of net transaction settlement (i.e., the simultaneous settlement of transactions in batches), which is addressed in the Night-time Settlement period. 
In this section, we outline its main components and highlight the main challenges faced by T2S. For a more detailed overview of the topic, we refer the reader to Ref.~\cite{alekseeva2020securities,T2SUDFS}.

\paragraph{Parties Accounts}
Each T2S user has at least one securities account (SAC) holding securities and one cash account (DCA) that holds cash. Each SAC and DCA are opened in the accounting books of CSDs and central banks, respectively. Moreover, each DCA is associated with a cash balance, and each SAC with one or more security positions. Excluding the one pertaining to SACs of security issuers or DCAs of central banks, each security position and cash balance cannot have a negative balance in their respective units.

\paragraph{Transactions}
A transaction is an agreement to exchange cash and/or securities between different cash balances and/or securities positions. Depending on the nature of the exchanges involved, transactions in T2S can be one of the following types: \emph{Delivery vs Payment} [DvP: exchange of securities against cash, as in Fig.~\ref{fig:cycles_dvp}(b)], \emph{Free of Payment} (FoP: delivery of securities), \emph{Payment Free of Delivery} (PfoD: delivery of cash), and \emph{Delivery with Payment} (DwP: delivery of cash and securities together). For each transaction, T2S defines a single level of priority based on the highest priority value among those set by the involved T2S actor (or automatically assigned by T2S) and the time elapsed from its intended settlement date \cite{T2Salgoobj}. Transactions can be linked together either via a \emph{with link} (requiring simultaneous settlement), or an \emph{after link} (where one must settle before or at the same time of the other).

\paragraph{Technical netting}

To minimize the resources required for settlement and prevent gridlocks \cite{bruno2022quantum}, T2S employs technical netting. Specifically, during the night-time settlement period, transactions are processed in batches, with settlement occurring on an all-or-none basis. This means that, during provision check, cash and securities necessary for settlement are calculated based on the net quantities of assets moved by all the settleable transactions [see Eqs.~\eqref{eq:nnb}]. By considering the net flows of exchanged assets, instead of processing each transaction in sequence, the system can assess the effective resource availability and accurately determine the necessity for intra-day credit.

\paragraph{Auto-collateralization}
T2S offers central banks and payment banks the possibility to provide cash to credit consumers via an automated process secured with eligible collateral. This feature aims to cover the lack of cash in DCAs by automatically pledging securities from eligible SACs. 
T2S automatic credit provision can occur in two ways. In \emph{central bank collateralization}, the central bank provides additional cash to the party in need. In \emph{client collateralization}, the party requiring credit has access to additional cash lent by a financial institution (e.g., the payment bank). In both cases, the amount of securities pledged as collateral shall not lead to the violation of predefined credit limits specific to each account. T2S distinguishes between \emph{auto-collateralization on flow} where the credit provided can be secured using the very same securities that are being purchased, and \emph{auto-collateralization on stock} where the securities are already held by the credit consumer \cite{T2SAutocoll}.

\paragraph{Partial settlement}
This feature allows the optimizer to settle a fraction of the securities and cash associated with a transaction. It is triggered when a transaction cannot be fully settled due to insufficient resources in one of the involved asset accounts. Partial settlement can only be activated for DvP and FoP transactions that meet specific threshold criteria (such as minimum cash amounts and quantities) and whose instructions are all eligible for this feature.

\begin{table*}[ht!]
    \centering
    \footnotesize
    \begin{tabularx}{\textwidth}{@{}lXl@{}}
        \toprule
        \textbf{Assumption}            & \textbf{Description}                                                             & \textbf{Impact}                                  \\ \midrule
        No Partial Settlement      & We do not consider the partial settlement feature, as this is outside the scope of run $4$ in Fig.~\ref{fig:cycles_dvp}(a). As such, rather than a MIP, we solve a Linear Integer Programming problem \cite{papadimitriou1998combinatorial}       & High       \\
        Simplified Collateral      & We consider only on-flow collateral from central bank collateralization since it is the first and most used for auto-collateralization. We further assume that the PLEDGE collateralization procedure is in place, hence the securities used as collateral are locked in an account pledged to the credit provider and cannot be used to settle other transactions \cite{T2SUDFS}.          & High       \\
        Non-Shared Collateral &  Each security position in the SACs is targeted by at most one CMB-Security Position link.  & High\\
        Single Credit Provider      & As in T2S, each DCA is associated with only one credit provider.     & Medium       \\
        DvP, FoP, PfoD      &  We only consider transactions of type DvP, FoP, and PfoD (these are prominent in T2S batches). We do not explicitly model system-generated (DvP-like) collateral transactions and other minor transaction types.   & Medium       \\
        Single Currency      & We consider only euro-related transactions. To generalize the model, one can pre-convert each currency amount to euros.     & Low       \\
        Link-After Only      & We consider only the transactional-links {after}. Links \emph{with} can always be defined from pairs of links \emph{after}.     & Low       \\
        Clean Data        & Cross-relations between input data are corrected and validated.                  & Low \\ \bottomrule
    \end{tabularx}
    \caption{
    Model's assumptions and impact on model generality. Notation and definitions of quantities in this table are given in Sec.~\ref{sec:NTSP_overview}. The impact is a qualitative indication of how easily these assumptions could be lifted in future generalization of our algorithm in Sec.~\ref{sec:methodology} (``Low'' $\rightarrow$ easy; ``High'' $\rightarrow$ hard).
    }
    \label{tab:assumptions}
\end{table*}

\subsection {Problem Modeling}
\label{sec:model}

\begin{figure*}
 \center
  \includegraphics[width=\textwidth]{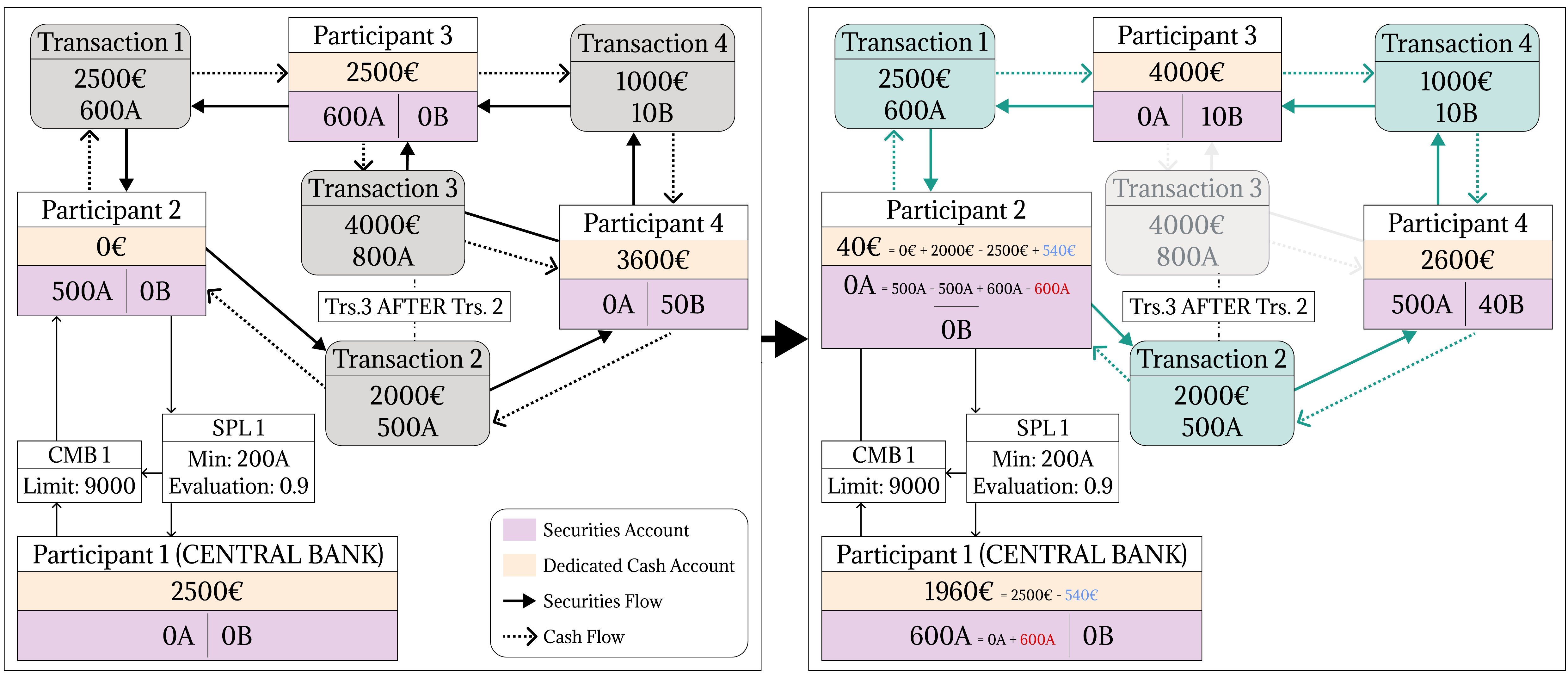}
  \caption{
  Example of $4$ transactions involving $3$ parties, exchanging securities A and B. Participant $2$'s security account is eligible for CoF. The limits for auto-collateralization ($\aLi_{\rm CMB 1} = 9000$\euro{}), as well as the type and value of pledgeable securities ($\qMi_{\rm SPL 1}=200$A, $\val_{A}=0.9$\euro{}/A), are determined by CMB $1$ and SPL $1$ (white boxes). On the left, we represent transactions as gray nodes connecting SAC (colored box, violet background) and DCA (colored box, orange background) of participants involved in the trades. Assets' movements are represented by means of arrows (dashed for cash, continuous for securities). An after-link is defined between transactions $2$ and $3$ ($3$ to be settle after/with $2$). On the right, we highlight in green the transactions that can be settled without violating constraints in Eqs.~\eqref{eq:nnb}-\eqref{eq:after_link}. Transaction $3$ cannot be settled due to insufficient securities of type A in participant $4$'s SAC. To demonstrate the effect of auto-collateralization, we explicitly compute the final balances for Participant $2$ and its creditor (Participant $1$). The final solution requires the collateralization of $600$ securities of type A from transaction $1$, generating $540$ \euro{} ($=600\text{A}\times0.9\text{\euro{}/A}$) in credit.
  }
  \label{fig:example_coll}
\end{figure*}

We present a simplified, yet representative, model of the NTSP, drawing on \cite{alekseeva2020securities} and the T2S functional documentation \cite{T2SUDFS}. We focus on one of the multiple optimization runs performed during the night-time cycle, specifically run $4$. As depicted in Fig.~\ref{fig:cycles_dvp}(a), each of these runs is a standalone optimization that processes run-specific transactions along with unsettled ones from previous runs. Run $4$ 
considers all types of transaction and is 
the most computationally demanding phase of 
the night-time cycle.

As explained in Table~\ref{tab:assumptions}, we model phase R4 of the NTSP (see Fig.~\ref{fig:cycles_dvp}) as a Linear Integer Programming problem, with decision variables divided into two groups. The binary vector $\X \in \{0, 1\}^{|\T|}$ is the settlement status of each transaction ($0$ for unsettled, $1$ for settled), and the integer vector $\Y \in {\mathbb{N}}^{|\SPL|}$ indicates the number of lots of securities pledged through auto-collateralization (see Sec.~\ref{sec:NTSP_overview} and \textbf{Auto-Collateralization} paragraphs below). Here, $|\T|$ is the number of transactions in the batch $\T$, and $|\SPL|$ is the number of Cash Management Bill (CMB)-Security position links that are required for auto-collateralization (see \textbf{Auto-Collateralization}). The optimizer's task is to find the optimal $\X$ and $\Y$ such that the objective function in Eq.~\eqref{eq:objf} is maximized while satisfying the constraints in Eqs.~\eqref{eq:nnb}-\eqref{eq:after_link}.  
The proposed model is built upon the assumptions listed in Table~\ref{tab:assumptions}. 

In Fig.~\ref{fig:example_coll}, we present a graphical example showcasing the security transaction settlement optimization process described in this section. We consider a simple NTSP instance of $4$ transactions which include all the elements discussed in the following paragraphs.

\paragraph{Objective Function:}
The role of the objective function $\ofun$ is to guide the optimization process towards the optimal solution of the considered NTSP instance. 
Following Ref.~\cite{T2Salgoobj}, for a set $\T$ of transactions and solutions $\X = \{ x_{1},\dots,x_{|\T|} \} \in \{0,1\}^{|\T|}$ 
we define $\ofun$ depending on total cash exchanged, volume, and priorities of the settled transactions. These terms are combined to yield

 \begin{equation}
 \label{eq:objf}
         \ofun(\X) = \lambda \dfrac{\sum_{\transa \in \T} \w_\transa \at \xt}{\sum_{\transa \in \T} 
         \w_\transa \at} + (1-\lambda) \dfrac{\sum_{\transa \in \T} \w_\transa \xt}{\sum_{\transa 
         \in \T} \w_\transa},
 \end{equation}
where $\at \in \mathbb{R}^{+}_{0}$ is the amount of cash exchanged with transaction $\transa$ and $\xt \in \{0,1\}$ indicates whether or not $\transa$ is settled. The values $\w_\transa \in \mathbb{R}$ are the user specified transaction's priority weights (see Sec.~\ref{sec:NTSP_overview}). 

The first term in Eq.~\eqref{eq:objf} is the (weighted) volume of settled cash, while the second terms is the (weighted) number of settled transactions. The free parameter $\lambda \in [0, 1]$ allows to vary the importance of volume over cash amounts. We remark that the decision variables $\Y \in {\mathbb{N}}^{|\SPL|}$ only indirectly contribute to the objective function in Eq.~\eqref{eq:objf} via the constraints described in \textbf{Auto-Collateralization} below.

\paragraph{Non-Negativity of Cash \& Securities Balances:}

\label{subsec:balance}
The objective function $\ofun(\X)$ in Eq.~\eqref{eq:objf} yields its maximal value $\ofun(\X) = 1$ when all transactions are settled, i.e., $\X = \{ 1,\dots,1 \}$. The complexity of the NTSP arises from the constraints that candidate solutions $\X$ must satisfy. The first of these constraints are the non-negativity of Cash and Securities Balances, and ensure that the assets leaving either cash balances or security positions are less than the available amounts. They are not applied for cash balances of central banks and security positions of security-issuers, and are modeled as follows:
\begin{subequations}
\label{eq:nnb}
    \begin{align}
        & \sum_{\transa \in \T_\cbal^{\debtor}} \at \xt - 
        \sum_{\transa \in \T_\cbal^{\creditor}} \at \xt
        \leq \aIn_\cbal + \aCo_\cbal ,\label{eq:nnb1}
        \\
        & \sum_{\transa \in \T_\spos^{\debtor}} \qt \xt - 
        \sum_{\transa \in \T_\spos^{\creditor}} \qt \xt
        \leq \qIn_\spos - \qCo_\spos 
        .\label{eq:nnb2}
    \end{align}
\end{subequations}
In Eq.~\eqref{eq:nnb1}, $\aIn_\cbal$ is the cash initially stored in the cash balance $\cbal$, and $\aCo_\cbal$ is the additional cash provided through auto-collateralization -- see below.
The sum $\sum_{\transa \in \T_\cbal^{\debtor}} \at \xt$ ($\sum_{\transa \in \T_\cbal^{\creditor}} \at \xt$) is the amount of cash that debits (credits) $\cbal$, with $\T_\cbal^{\debtor}$ ($\T_\cbal^\creditor$) being the set of transactions debiting (crediting) $\cbal$. Overall, Eq.~\eqref{eq:nnb1} ensures that the net outflow of cash from $\cbal$ is not greater than the available amount. 

Similarly, Eq.~\eqref{eq:nnb2} checks that the net quantity of securities leaving a security position $\spos$ is not greater than $\qIn_\spos$ (the initial quantity in $\spos$) minus $\qCo_\spos$ (the quantity in $\spos$ pledged as collateral). Here, $\qt$ is the quantity of securities exchanged with transaction $\transa$, and the two summations represent the quantity of securities that, in the order, debits and credits the security position $\spos$.

\paragraph{Auto-Collateralization:}
\label{subsec:coll}
Given the complexity of the auto-collateralization mechanism \cite{T2SAutocoll}, we model a simplified version which includes \emph{collateral on-flow} (CoF) processes, see Table~\ref{tab:assumptions}. Furthermore, we only consider \emph{primary} collateralization, where the credit-lender is a central bank, not subject to balance restrictions. 

The CoF mechanism could be triggered if a cash balance $b_\transa^\debtor$ cannot cover for the cash amount required by a transaction $\transa$. When this happens, part of the incoming securities are pledged to mitigate the cash shortfall. Not all transactions can access auto-collateralization; their eligibility is regulated by the credit memorandum balances in $\CMB$ and the CMB-security position links in $\SPL$. The first is a 
set whose elements $\cmb \in \CMB$ establish a link between a cash balance that can enjoy CoF with the cash balance of a central bank (see CMB 1 in Fig.~\ref{fig:example_coll}, connecting the cash balance of participant 2 with that of a central bank, i.e. participant 1). For CoF to take place, $\cbal_\transa^\debtor$ must be registered in $\CMB$.

On the other hand, each element $l \in \SPL$ connects a security position $\spos_\spl$ with a $\cmb_\spl \in \CMB$ and its associated cash balance $\cbal_\cmb$ that may request credit (see SPL 1 in Fig.~\ref{fig:example_coll}, connecting the security positions A of the SAC of participant 2 with that of a central bank -- where securities will be stored if pledged -- and CMB 1). In other words, for a transaction $t$ to be eligible for auto-collateralization, not only a paying cash balance $b_\transa^\debtor$ has to be linked to a central bank via a $\cmb \in \CMB$. The receiving security position $\spos^\creditor_\transa$ involved in the exchange must also be authorized to trigger the CoF mechanism via a link 
\begin{equation*}
    \spl \in \SPL \text{ such that }
    \begin{cases}
        (\cmb = \cmb_\spl) \land (b_\transa^\debtor = \cbal_\cmb)
        \\
        \spos^\creditor_\transa = \spos_\spl
    \end{cases}
    .
\end{equation*}
Here, $\cmb_\spl$ is the $\CMB$ associated with $\spl$ and $\cbal_\cmb$ is the cash balance receiving the collateralized-cash.

For each $\spl \in \SPL$, we must calculate the number of lots $\yp \in \mathbb{N}$ of security $\secu_{\spos_\spl}$ pledged from the security position $\spos_\spl$ through $\spl$.  
The quantities [see Eq.~\eqref{eq:nnb}]
\begin{subequations}
    \begin{align*}
        \qCo_\spos
        & =
        \sum_{\spl \in \SPL_\spos} \lot_{\secu_\spos} \yp
        ,\\
        \aCo_\cbal
        & =
        \sum_{\spl \in \SPL_\cbal} \qCo_\spl \val_{\secu_\spl}
        ,
    \end{align*}
\end{subequations}
are the total number of collateralized securities from each security position $\spos$, and the total amount of credit added to each cash balance $\cbal$, respectively. 
Here, $\SPL_\spos$ ($\SPL_\cbal$) are all links $\spl \in \SPL$ that enable auto-collateralization for a security position $\spos$ (cash balance $\cbal$). Furthermore, $\lot_{\secu_\spos}$ is the size of a single lot of security of type $\secu_\spos$, $\val_{\secu_\spl}$ is the daily valuation in euros of a unit of security $\secu_\spl$ (provided by central banks), and $\qCo_\spl = \lot_{\secu_\spl}\yp$.

Finally, cash and security positions exchanged within the auto-collateralization mechanism are constrained to not exceed predefined limits. For all $\cbal$ ($\spl \in \SPL$) such that $\exists~ \cmb \in \CMB:~ \cbal= \cbal_\cmb$ ($\qCo_\spl \neq 0$)
\begin{subequations}
\label{eq:cash_sec_coll}
    \begin{align}
        \aNe_\cbal 
        &\leq 
        \aCo_\cbal \leq \aLi_\cmb
        , \label{eq:cash_coll}
        \\
        \qMi_\spl 
        &\leq 
        \qCo_\spl \leq \qLi_\spl
        , \label{eq:sec_coll}
    \end{align}
\end{subequations}
where 
\begin{equation*}
    \aNe_\cbal = \max \left\{ 0, ~
        \sum_{\transa \in \T_\cbal^\debtor} \at \xt - 
        \sum_{\transa \in \T_\cbal^\creditor} \at \xt - 
        \aIn_\cbal \right\}
\end{equation*}
is the amount of additional credit required to settle all the transactions associated to a cash balance $\cbal$. In turn, $\aNe_\cbal$ allows us to determine 
\begin{equation*}
\label{eq:auc1}
        \qLi_\spl = 
        \begin{cases}
            \sum_{\transa \in \T_\spl} \xt\qt & \text{if } \aNe_\cbal > 0 
            \\
            0 & \text{otherwise}
        \end{cases}
        ,
\end{equation*}
i.e., the maximum number of securities that can be exchanged through the link $\spl$. Here, $\T_\spl$ is the set of transactions that can activate CoF through $\spl$.

Eq.~\eqref{eq:cash_coll} ensures that the amount of credit added to the cash balance $\cbal_\cmb$ is greater than $\aNe_\cbal$ (the required additional cash to carry out the transaction) and lower than the credit limit $\aLi_\cmb$ set by $\cmb \in \CMB$.        
On the other hand, Eq.~\eqref{eq:sec_coll} ensures that when the auto-collateralization mechanism is activated ($\qCo_\spl \neq 0$), the quantity $\qCo_\spl$ of collateral pledged in the context of an $\spl \in \SPL$ lies within the range $[\qMi_\spl , \qLi_\spl]$.    

\paragraph{After Links:}
These are associations between two or more transactions to be processed. An after link allows a transaction $\transa_2$ to be settled only after (or at the same time) another specified transaction $\transa_1$ has been settled (see, for instance, transactions 2 and 3 in Fig.~\ref{fig:example_coll}).
We model the after link by the constraint
\begin{equation}
\label{eq:after_link}
    \x_{\transa_2} \leq \x_{\transa_1},
\end{equation}
where the inequality impedes the settlement of $\transa_2$ ($\x_{\transa_2} = 1$) unless $\x_{\transa_1} = 1$, i.e., $\transa_1$ is also settled.

\begin{figure*}[!t]
\center
\includegraphics[width=\textwidth]{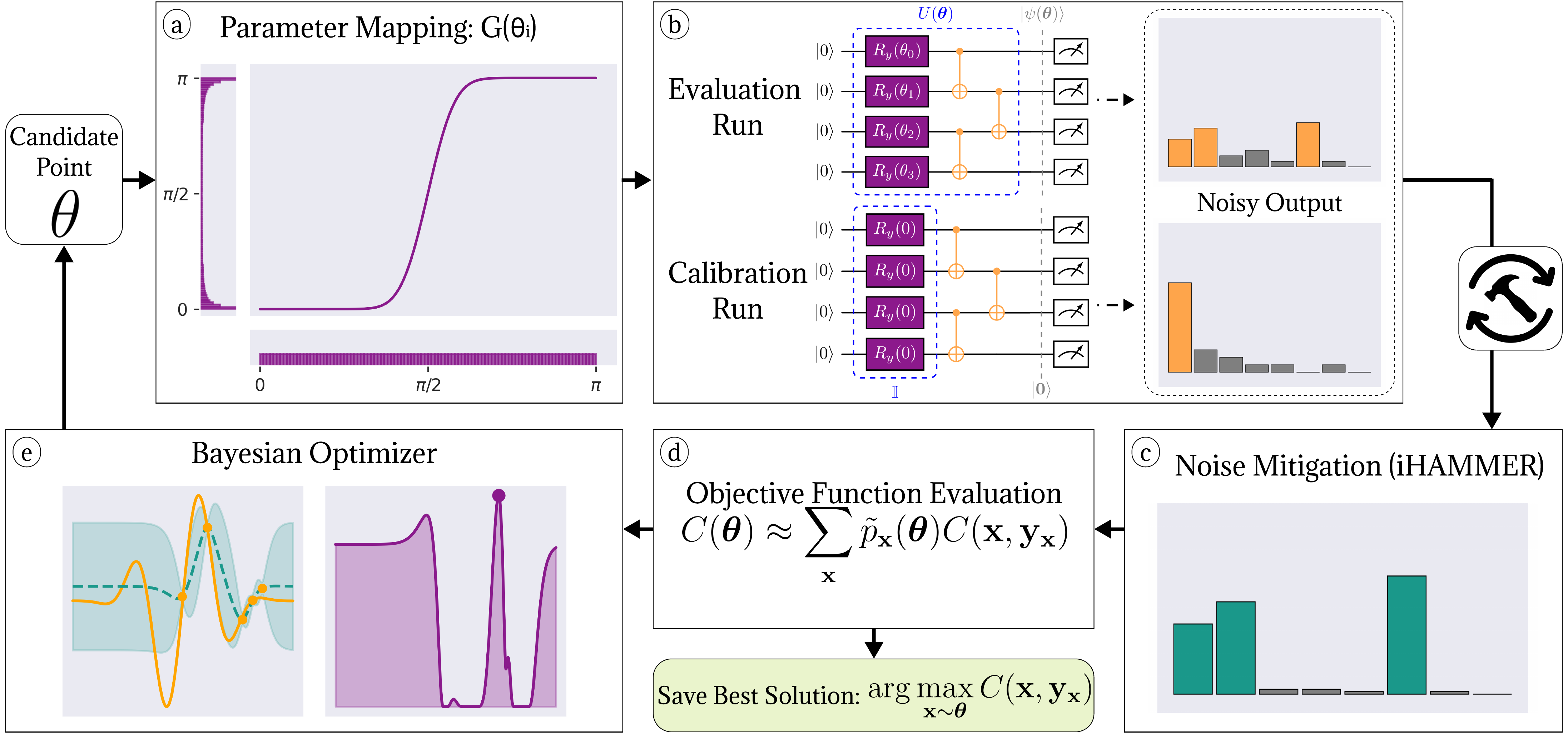}
\caption{
Architecture of the proposed VQA. (a) Starting from a point $\boldsymbol{\hat{\theta}}$ in the parameters space, we apply the parameter mapping function in Eq.~\eqref{eq:mapping} to obtain the ansatz parameters $\boldsymbol{\theta}$. (b) We run an evaluation and a calibration quantum circuits to prepare and measure the states $\ket{\psi(\boldsymbol{\theta})} = U(\boldsymbol{\theta})\ket{0}^{\otimes n}$ and $\ket{\psi(\boldsymbol{0})} = \ket{0}^{\otimes n}$, respectively. (c) From the noisy outputs obtained in the previous step, iHAMMER reconstructs the probability $p_{\X}(\boldsymbol{\theta})$ of measuring the candidate solutions $\X$ from $\ket{\psi(\boldsymbol{\theta})}$. (d) These corrected probabilities are used to compute the cost function $C(\boldsymbol{\theta})$ in Eq.~\eqref{eq:final_obj}, and we save the best solution $\X$ obtained in the measurement process (and the associated vector $\Y_\X$ calculated using Algorithm~\ref{algo:greedy_coll}). (e) Finally, BO (see App.~\ref{sec:app_bayesian_opt}) updates the parameters and starts a new iteration.
}
\label{fig:algo}
\end{figure*}

\section{Methodology}
\label{sec:methodology}

Variational quantum algorithms (VQAs) are hybrid quantum-classical algorithms that are targeted to NISQ \cite{preskill2018quantum,callison2022} devices and have been employed to address a range of problems, from chemistry \cite{motta2022emerging, kandala_error_2019,chan2024measurement}  and physics \cite{banuls2020simulating, daley2022practical, kokail2019self, paulson2020qed, haase2021qed, ferguson2021measurement} to optimization \cite{farhi_quantum_2022, Zhu_VQA_x_OPT, Tabi_VQA_x_OPT} and machine learning \cite{cerezo2021variational,williams2024addressing,umeano2024geometric,scali2024quantum,bowles2024better}.
The high interest in VQAs stems from their ability to mitigate hardware noise by exploiting shallow quantum circuits.

Quantum approaches addressing the NTSP have been discussed in Refs.~\cite{braine2021quantum,huber2024exponential}.
Both consider a simplified version of the problem introduced in Secs.~\ref{sec:STSO} and \ref{sec:methodology}. The algorithm in Ref.~\cite{braine2021quantum} is tested on simplified instances of the NTSP with up to $7$ transactions. 
In Ref.~\cite{huber2024exponential}, on the other hand, the qubit-efficient encoding from \cite{tan2021qubit} encodes the solution $\mathbf{x}$ of the problem into $M=\lceil \log(n) \rceil + 1$ qubits (instead of $n$), where $n = |\T|$ is the dimension of $\mathbf{x}$. Although this reduction in qubit number allows addressing larger problem instances of the NTSP, it comes with limitations. Specifically, reducing the number of qubits increases the sampling overhead, requiring more measurements to reconstruct a complete solution.

We adapt the VQA architecture to the NTSP as defined in Sec.~\ref{sec:model}, incorporating constraints as penalty terms in the cost function and adopting a sampling-based approach for the evaluation of the quantum circuit's output. 
The final custom optimization pipeline is depicted in Fig.~\ref{fig:algo}, and its key components are discussed below.

\paragraph{Parameterized Ansatz [Fig.~\ref{fig:algo} (b)]:}

The parameterized quantum circuit $U(\boldsymbol{\theta})$ used in this study follows the structure of the hardware-efficient ansatz illustrated in Fig.~\ref{fig:algo}(b). It consists of $N_{\rm L}$ repetitions of $3$ layers of gates ($N_{\rm L} = 1$ in the figure) applied to a register of $n = |\T|$ qubits initialized in $\ket{0}^{\otimes n}$. The first layer is a set of parameterized single-qubit rotations $R_y(\theta)$ and the following two are made of C-NOT gates acting on nearest neighbors. At the end of the circuit, each qubit is measured, yielding a value in $\{0,1\}$ with a probability dependent on $\boldsymbol{\theta}$, i.e., all parameters within the $N_{\rm L}$ rotation layers.

The width of the ansatz is the number of transactions \( n=|\T| \) in the considered NTSP, while its depth is $N_{\rm L}+2N_{\rm L}$ (rotations' $+$ C-NOTs' layers), i.e. $3N_{\rm L}$.
We have chosen this ansatz to match the connectivity of IBM devices \cite{IBM2024roadmap} -- hence its ``hardware-efficient'' nature (SWAPs are not required in the transpilation process \cite{HardwareEff}). We remark that other types of ansatz -- particularly the ones performing adiabatic shortcuts \cite{zhou2020quantum} -- are more promising to showcase a quantum advantage (see Sec.~\ref{sec:discussion}). However, they generally require depths and numbers of entangling gates that are prohibitive for current devices.

\paragraph{Parameters Mapping [Fig.~\ref{fig:algo} (a)]:}

To mitigate the effect of statistical errors in estimating $\tilde{p}_{\mathbf{x}}$ in Eq.~\eqref{eq:final_obj}, we implement a parameter mapping for each $\theta_i \in \boldsymbol{\theta}$. The idea is to facilitate the quantum state $\ket{\psi (\boldsymbol{\theta})} = U(\boldsymbol{\theta}) \ket{0}$ to be localized, such that $\tilde{p}_{\mathbf{x}}(\boldsymbol{\theta}) \approx p_{\mathbf{x}}(\boldsymbol{\theta})$ even for large $n$ and a polynomial number of shots (i.e. quantum circuit measurements). 
As shown in Fig.~\ref{fig:algo}, this parameter mapping is a function $G(\theta_{i})$ that modifies the input parameters $\theta_{i} \in \boldsymbol{\theta}$ by shifting the rotation angles within $R_y(\theta_{i})$ towards the closest value between $0$ and $\pi$. This promotes a sparse output distribution.

We build the mapping $G: [0, \pi] \mapsto [0, \pi]$ from Gaussian functions $g_{\mu, \sigma}$ centered in $\mu$ and with standard deviation $\sigma$, 
\begin{equation}
\label{eq:mapping}
    \text{G}(\theta_i) = \frac{\sum_{\mu\in \{0, \pi\} }\int_0^{\theta_i+\frac{\pi}{2}}g_{\mu, \sigma}(x)~dx}{\frac{2}{\pi} \int_0^\pi g_{0, \sigma}(x)~dx} - \frac{\pi}{2}
    ,
\end{equation}
for each $\theta_i \in \boldsymbol{\theta}$.
The standard deviation $\sigma$ acts as a hyperparameter, allowing for fine-tuning of the function's behavior. It can be regulated based on system size and the desired level of sparsity -- the values employed in this work are reported in App.~\ref{sec:app_conf}.

\paragraph{Model Encoding Through Activation Functions [Fig.~\ref{fig:algo} (d)]:}

We introduce a non-linear parameterized objective function to model the NTSP formalized in Sec.~\ref{sec:model}. The core idea involves adding the $V$ constraints $g_v(\mathbf{x}, \mathbf{y}) + b_v \leq 0$ obtained from Eqs.~\eqref{eq:nnb}-\eqref{eq:after_link} as non-linear penalty terms to the payoff function in Eq.~\eqref{eq:objf}. Here, the number $V$, the functions $g_v$ and the constants $b_v$ depend on the problem at hand, as described in Sec.~\ref{sec:model}.
More formally, given an NTSP instance with $n = |\T|$ transactions, we define the associated unconstrained objective function as:
\begin{equation}
     C(\mathbf{x}, \mathbf{y}) = \ofun(\mathbf{x}) + \sum_{v = 1}^{V} h_{v}(g_{v}(\mathbf{x}, \mathbf{y}) + b_v),
\label{eq:activ_obj_fun}
\end{equation}
where $\ofun(\mathbf{x})$ is the payoff in Eq.~\eqref{eq:objf} and $h_{v}$ are non-linear activation functions acting on the NTSP constraints (in canonical form) $g_v(\mathbf{x}, \mathbf{y}) + b_v \leq 0$. The activation functions $h_{v}$ are chosen to penalize the values of $\mathbf{x}\in\{0,1\}^n$ and $\mathbf{y}\in\mathbb{N}^j$ when $g_{v}(\mathbf{x}, \mathbf{y}) + b_v > 0$, thus favoring solutions where the variables satisfy all the constraints in Eqs.~\eqref{eq:nnb}-\eqref{eq:after_link}. Here, $j$ is the number of CMB-security position links in the NTSP. For a discussion about the $h_{v}$ chosen in this work, see App.~\ref{sec:app_obj_fun_conf}.

Due to the non-linearity of the activation functions $h_{v}$ in Eq.~\eqref{eq:activ_obj_fun}, $C(\mathbf{x}, \mathbf{y})$ cannot be easily mapped into a weighted sum of Pauli operators that are routinely measured on quantum computers \cite{shlosberg2023adaptive}. However, Eq.~\eqref{eq:activ_obj_fun} is a diagonal Ising-type observable \cite{bravyi2017complexity}, meaning that each quantum state $\ket{\mathbf{x}}$ encoding $\mathbf{x}$ in the computational basis is a candidate solution to the problem. For any $\mathbf{y}$, it is therefore classically efficient to find the associated $C(\mathbf{x}, \mathbf{y})$ in Eq.~\eqref{eq:activ_obj_fun}. As such, from a set of input parameters $\boldsymbol{\theta}$ (see below and in Fig.~\ref{fig:algo}), we can estimate the average cost of the quantum state $\ket{\psi(\boldsymbol{\theta})}$ by
\begin{equation}
    \label{eq:final_obj}
    C(\boldsymbol{\theta}) \approx \sum_{\mathbf{x}} \tilde{p}_{\mathbf{x}}(\boldsymbol{\theta}) C(\mathbf{x}, \mathbf{y_{\mathbf{x}}}),
\end{equation}
where $\Y_\X$ is efficiently computed from $\X$ with the heuristic in Algorithm~\ref{algo:greedy_coll}. Furthermore, $\tilde{p}_{\mathbf{x}}(\boldsymbol{\theta})$ is the reconstructed probability (from repeated measurements) of obtaining $\mathbf{x}$ from $\ket{\psi(\boldsymbol{\theta})}$, i.e., $\tilde{p}_{\mathbf{x}}(\boldsymbol{\theta}) \approx p_{\mathbf{x}}(\boldsymbol{\theta}) = \lvert \langle \mathbf{x} \vert \psi(\boldsymbol{\theta})\rangle \rvert^2 $.

In summary, Eq.~\eqref{eq:final_obj} is computed from the individual measurement output provided by the quantum computer, rather than from the estimated averages of Pauli operators \cite{shlosberg2023adaptive}. While this allows to incorporate nonlinear functionals in $C(\boldsymbol{\theta})$, this paradigm shift has a considerable impact on the error mitigation techniques that can be employed (see Sec.~\ref{sec:discussion}).

\paragraph{Noise Mitigation [Fig.~\ref{fig:algo} (c)]:}

Error mitigation \cite{chan2024measurement,rahman2022self_error_mitigation, urbanek2021_error_mitigation, miguelramiro2023enhancing,miguelramiro2023sqem_error_mitigation} is crucial to execute algorithms on NISQ hardware \cite{preskill2018quantum}. 
In this work, we use dynamical decoupling (DD) \cite{viola1999dynamical}, and a modified version (named iHAMMER) of the Hamming Reconstruction technique first presented in Ref.~\cite{tannu2022hammer}. 
In the experiments discussed in Sec.~\ref{sec:experiment}, DD is included at the moment of circuit transpilation leveraging Qiskit integrated subroutines \cite{javadi2024quantum}. These methods slightly modify the structure of the ansatz implementing digital DD as discussed in \cite{jurcevic2021demonstration}.  
iHAMMER, on the other hand, modifies the reconstructed probabilities $\tilde{p}_{\mathbf{x}}(\boldsymbol{\theta})$  in Eq.~\eqref{eq:final_obj} following a physically motivated heuristic. It relies on empirical evidence that errors tend to be ``localized'' nearby the correct state 
to enhance the probability of outcomes having ``rich neighborhoods'' and high sampling counts, while reducing the likelihood of isolated bitstrings. Compared to the original work in \cite{tannu2022hammer}, iHAMMER is an iterative procedure that continues until a threshold is met. This threshold aims at preserving the structure of the desired quantum state, and is calculated from a calibration routine that estimates the number of errors that occur with the specific circuits employed by the VQA, see Fig.~\ref{fig:algo}(b). For more details on iHAMMER, see App.~\ref{sec:app_ihammer}.

\paragraph{Classical Optimizer [Fig.~\ref{fig:algo} (e)]:}

In this work we employ Bayesian Optimization (BO) \cite{BO} as the classical optimizer for the VQA. BO is a sequential optimization strategy for black-box functions, particularly effective for non-convex functions that are expensive to evaluate and potentially noisy. It relies on a probabilistic model that approximates the objective function [in our case Eq.~\eqref{eq:final_obj}], and an acquisition function that guides the search process based on the probabilistic model.

BO has two main advantages. The first is its sample efficiency, requiring fewer function evaluations compared to other global optimization methods \cite{Bayesian_efficient}. Second, it naturally incorporates noise in observations and does not require gradient information \cite{Bayesian_no_gradient}. As a downside, BO's probabilistic model scales as $\mathcal{O}(i^3)$ (where $i$ is the number of objective function evaluation), which is larger than other approaches \cite{venter2010review}. Furthermore, despite recent advances \cite{SAASBO}, BO seems to suffer in high dimensional spaces due to the complexity in approximating the objective function. Further details on BO are discussed in App.~\ref{sec:app_bayesian_opt}.

\section{Experiments}
\label{sec:experiment}

In this section, we present the benchmark of our VQA on IBM superconducting-qubits quantum devices \cite{roadmap-ibm}. We use Qiskit SDK \cite{javadi2024quantum} for compiling the quantum circuits and to interface with the quantum computers. 
We discuss in Sec.~\ref{sec:noise_mit_eval}  the effects of the iHAMMER technique for error mitigation on sparse quantum states.  Sec.~\ref{sec:vqa_performance} is focused on the VQA benchmarks, while in Sec.~\ref{sec:discussion} we address limitations and possible improvements of the tested VQA.

\subsection{Noise Mitigation}
\label{sec:noise_mit_eval}

First, we analyze the performance of our error mitigation technique (see also App.~\ref{sec:app_ihammer}) by comparing the outputs of real quantum hardware corrected and non-corrected by iHAMMER. Following the parameter mapping described in Sec.~\ref{sec:methodology}, we consider sparse quantum states that are similar to the ones typically encountered in our VQA optimization process. This is essential to correctly estimate the probabilities $\tilde{p}_{\mathbf{x}}(\boldsymbol{\theta}) \approx p_{\mathbf{x}}(\boldsymbol{\theta})$ within the objective function in Eq.~\eqref{eq:final_obj}.

Following the ansatz described in Sec.~\ref{sec:methodology} and Fig.~\ref{fig:algo}(b), we choose the parameterized quantum circuits with $N_{\rm L} = 1$ and $n$ ranging from $25$ to $50$ qubits. Eight of the parameters $\boldsymbol{\theta}$ (randomly chosen) are set to $\frac{\pi}{3}$, while all others are kept to $0$. This way, $p_{\mathbf{x}}(\boldsymbol{\theta})$ 
always contains $2^8$ peaks of known heights, independently from $n$. 

\begin{figure}[ht]
    \centering
    \includegraphics[width=\linewidth]{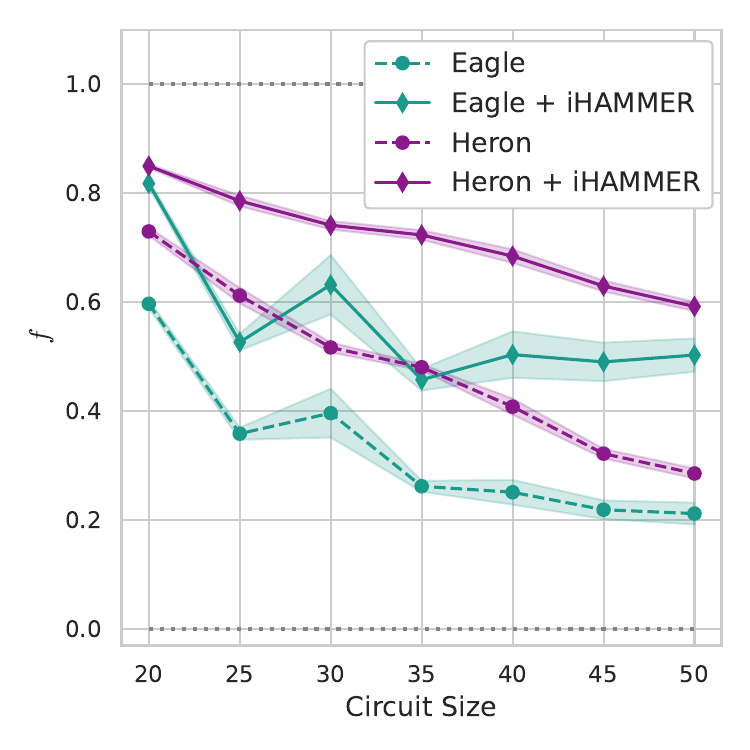}
    \caption{
    Comparison between fidelity of sparse quantum states obtained before and after (+ iHAMMER) noise mitigation (see main text for more details). Points are averages of $10$ runs and shadowed areas represent their standard deviations.
    }
    \label{fig:fidelity}
\end{figure}

In Fig.~\ref{fig:fidelity} we plot the fidelities $\textit{f}$ of the raw and mitigated output distributions with respect to the exact quantum state, $
\textit{f} 
= 
\left(\sum_{\mathbf{x}}
\sqrt{
\tilde{p}_{\mathbf{x}}(\boldsymbol{\theta}) \cdot p_{\mathbf{x}}(\boldsymbol{\theta})
}
\right)^{2} 
$. 
As reported in the legend, runs are executed on Heron (\texttt{ibm\_fez}, violet) and Eagle (\texttt{ibm\_brisbane}, green) QPUs, respectively the last and previous to last generations of IBM quantum devices (at the moment of writing). For each configuration, the points connected by lines are the average fidelities $\textit{f}$ computed over $10$ different runs, and shadows are their standard deviations.  

For both devices, fidelity decreases as circuit size increases, confirming that larger circuits are more subject to noise, which affects not only the raw outputs (dots connected by dashed lines) but also the ability of iHAMMER to identify and filter out noisy samples (diamonds connected by full lines). Nevertheless, the fidelity trend line for mitigated distributions consistently remains above that of the raw outputs, indicating that iHAMMER reliably enhances the accuracy of the sampling process. This enhancement is always substantial, consistent, and tends to improve at larger numbers of qubits.

In agreement with the findings from Ref.~\cite{mckay2023benchmarking}, we observe greater resilience to noise in the new generation of Heron QPUs available on \texttt{ibm\_fez}, compared to the Eagle QPUs used on \texttt{ibm\_brisbane}. 
The bump observed at $30$ qubits for \texttt{ibm\_brisbane} can be attributed to variability in noise conditions; $5$ out of the $10$ runs for this configuration were conducted at different times, during which the device experienced significantly different noise levels. 
Indeed, the operational conditions of IBM QPUs vary depending on several external factors, such as the time since the last calibration \cite{calibration}.

\subsection{VQA Performance Analysis}
\begin{figure*}[t]
    \includegraphics[width=\linewidth]{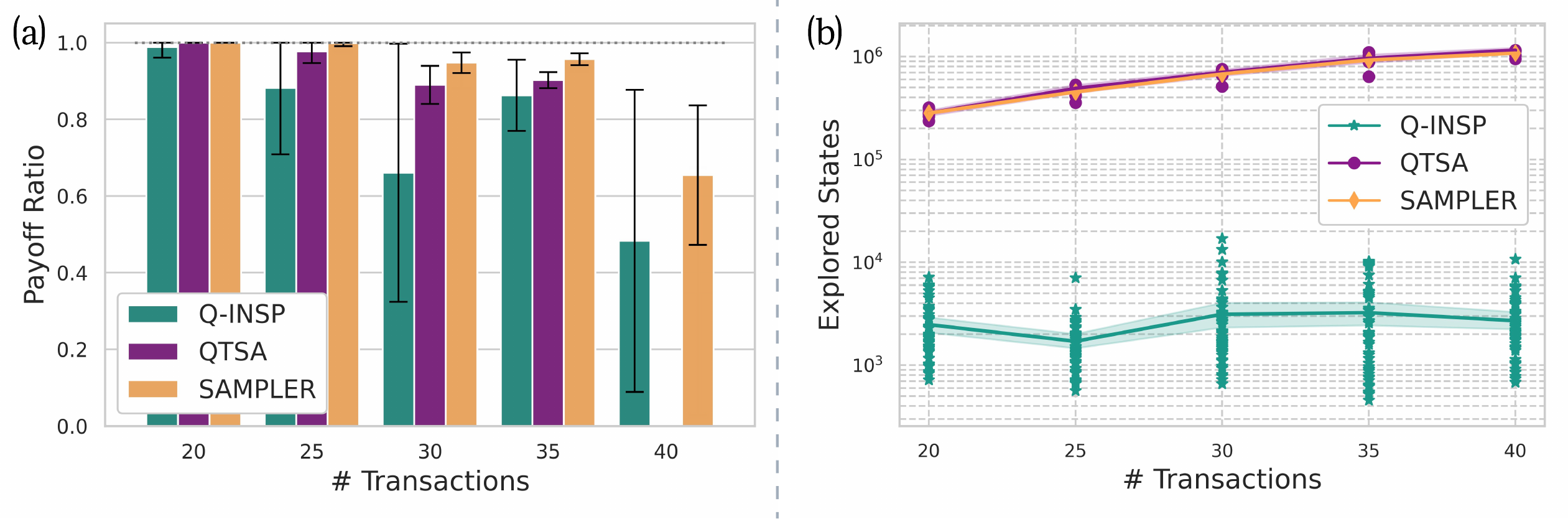}
    \caption{(a) Performance of our VQA approach in Sec.~\ref{sec:methodology} on NTSP instances with up to $40$ transactions. For each instance size, we plot the average payoff of the best solution obtained by running $5$ times QTSA (violet), $50$ times Q-INSP (green), and $50$ times SAMPLER (orange). Error bars indicate the payoff standard deviations. (b) Number of different solution (states) evaluated during the execution of the algorithms. Violet circles (green stars) [orange diamonds] refer to QTSA (Q-INSP) [SAMPLER]. Each line in the plot is the average, and the shadows represent the $95$\% confidence interval. For each NTSP instance size, SAMPLER evaluates a fixed number of solutions matching the average number of distinct samples processed by QTSA (see App.~\ref{sec:app_conf} for the exact values).}
    \label{fig:performance}
\end{figure*}
\label{sec:vqa_performance}
In this section, we benchmark the VQA solver introduced in Sec.~\ref{sec:methodology} on NTSP instances of $20$ to $40$ transactions. 
These real-life instances are sampled from the transactions that are processed by T2S in the night-time cycle and are subject to all real-world NTSP constraints discussed in Sec.~\ref{sec:model}. The sampling is made from anonymized transaction data provided by a European CSD, and is made such that the assumptions in Table~\ref{tab:assumptions} are satisfied. 
We restrict the benchmark on instances with at most $40$ transactions (i.e., qubits) due to the limitations affecting NISQ devices, both in terms of available qubits and coherence (see Sec.~\ref{sec:noise_mit_eval}).
The properties (size,traded securities, links, optimal payoff, percentage of transactions settled) of these instances are in Table~\ref{tab:data_stats}.
\begin{table}[ht!]
    \centering
    \footnotesize
    \begin{tabular}{lccccc}
    \toprule
    \textbf{Name} & \textbf{\#T} & \textbf{\#S} & \textbf{\#L} & \textbf{OP} & \textbf{PST} \\
    \midrule
    gs\_20 & 20 & 3 & 6 & 0.99997 & 70 \% \\
    gs\_25 & 25 & 3 & 8 & 0.80575 & 76 \% \\
    gs\_30 & 30 & 3 & 9 & 0.99998 & 66 \% \\
    gs\_35 & 35 & 3 & 11 & 0.82692 & 77 \% \\
    gs\_40 & 40 & 4 & 9 & 0.70324 & 60 \% \\
    \bottomrule
    \end{tabular}
    \caption{Instances statistics. \textbf{Name}: dataset name; \textbf{\#T}: number of transactions; \textbf{\#S}: number of different securities traded; \textbf{\#L}: number of after-links; \textbf{OP}: optimal payoff found by CPLEX; \textbf{PST}: percentage of transactions settled by the optimal solution.}
    \label{tab:data_stats}
\end{table}

To quantify the performance of the VQA, we compare our approaches with a CPLEX-based classical solver \cite{CPLEX} that always finds the optimal solution vectors $\X$ and $\Y$ and the corresponding payoff $\mathcal{F}(\X,\Y)$ in Eq.~\eqref{eq:objf}. In contrast with the VQA (which does not ensure optimality), this solver requires an exponential time, meaning that it can only be employed for small NTSP instances. 
In the following, we present our numerical results in terms of $\rho \in [0,1] $, i.e., the ratio between the output of the considered VQA algorithm and the CPLEX solution. Therefore, $\rho = 1$ [$\rho = 0$] indicates that the optimal solution [no solution respecting the constraints in Eqs.~\eqref{eq:nnb}-\eqref{eq:after_link}] is found.


As shown in Fig.~\ref{fig:performance}, we consider three distinct solvers, QTSA, Q-INSP and SAMPLER, applied to the instances in Table~\ref{tab:data_stats}.
QTSA (Quantum Transaction Settlement Algorithm) and Q-INSP (Quantum INSPired) are, respectively, a quantum and a quantum-inspired solvers based on the VQA pipeline in Sec.~\ref{sec:methodology}. The only difference between them is that Q-INSP is based on an unentangled ansatz with only a single layer of parametrized $R_y$ rotations ($U(\boldsymbol{\theta})$ in Fig.~\ref{fig:algo}(b) without C-NOTs). As a consequence, Q-INSP scales linearly with respect to the number of qubits classically and allows the evaluation of the optimization pipeline without hardware noise.
Finally, SAMPLER randomly evaluates a fixed number of different solutions $\X \in \{0,1\}^n$ and yields the one with the best objective function $C(\X,\Y_{\X})$ in Eq.~\eqref{eq:activ_obj_fun},where $\Y_{\X}$ is determined with Algorithm~\ref{algo:greedy_coll}.

For each NTSP instance, we execute $50$ runs with Q-INSP and SAMPLER, and $5$ different optimization attempts with QTSA (due to limited availability of the quantum devices). Specifically, QTSA runs were conducted on IBM's Eagle QPUs, using both \texttt{ibm\_brisbane} and \texttt{ibm\_sherbrooke} \cite{IBM2024roadmap}. Each run of QTSA and Q-INSP involves $300$ iterations (i.e. parameter updates), with $10^4$ samples taken per iteration to compute the probabilities $\tilde{p}_{\mathbf{x}}(\boldsymbol{\theta})$ in Eq.~\eqref{eq:final_obj} -- see Sec.~\ref{sec:methodology}. For every run of QTSA, iHAMMER has been employed to mitigate hardware noise.  The decision to fix solver resources regardless of instance size accounts for the fixed time constraints governing NTSP resolution in T2S \cite{T2SUDFS}; full configurations used for this experiment are detailed in App.~\ref{sec:app_conf}.

Fig.~\ref{fig:performance}(a) reports the average payoff ratio $\rho \in [0,1]$ for each solver and for each NTSP instance. The black error-lines on top of each bar are the standard deviations from repeated runs.
For all solvers, we observe a decline in $\rho$ as the input size increases. This is a consequence of the fixed resources (number of iterations and measurement shots) provided to each solver across different problem sizes, that penalizes larger instances.  
Although QTSA performs well for batches of $\leq 35$ transactions, Q-INSP and SAMPLER find valid solutions (that satisfy all constraints) for instances of $40$ transactions. 
Notably, the performance of QTSA and SAMPLER are similar across instances with $\leq 35$ transactions, both in terms of the average payoff ratios $\rho$ and standard deviations. As discussed in more details in Sec.~\ref{sec:discussion}, this suggests that (as a consequence of hardware noise) QTSA behaves similarly to a random sampler \cite{scriva2024challenges}.

\begin{figure*}[t!]
    \centering
        \includegraphics[width=\linewidth]{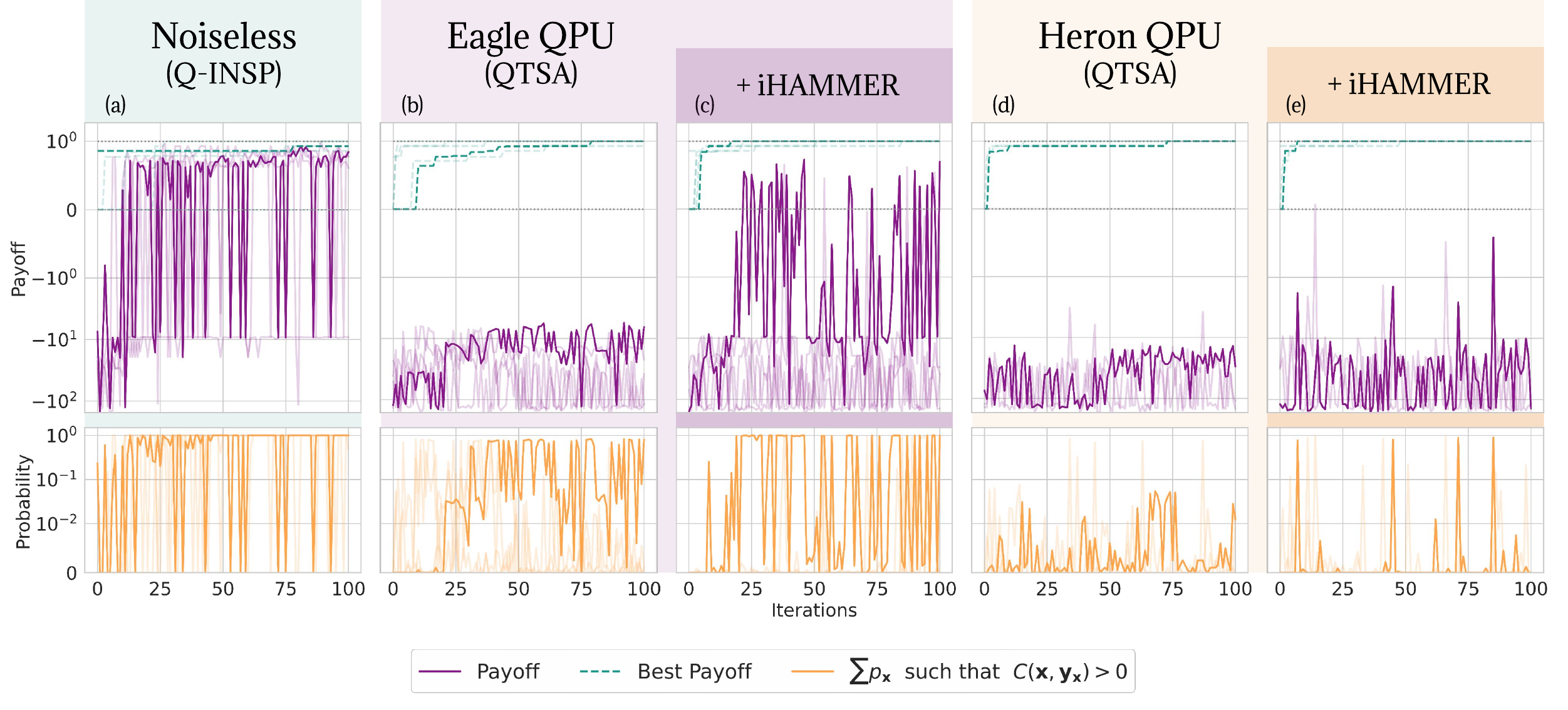}
    \caption{Payoff convergence analysis for QTSA and Q-INSP algorithms, solving gs\_20 (see Tab.~\ref{tab:data_stats}). (a) refers to Q-INSP noiseless runs executed on CPU, (b)[(c)] refers to QTSA runs executed on Eagle QPUs without [with] iHAMMER, and (d)[(e)] refers to QTSA runs executed on Heron QPUs without [with] iHAMMER. In each plot, the green dashed line represents the payoff of the best solution among the ones already sampled. The violet line is the payoff trend computed for each iteration as in Eq.~\eqref{eq:final_obj}. The cumulative probability $ \sum_\X \tilde{p}_\X $ of valid solutions sampled at each iteration is shown by the orange line. For clarity purpose, we highlight the trend lines of the runs that most closely match the expected convergence behavior of BO (see main text).}
    \label{fig:convergence}
\end{figure*}

In Fig.~\ref{fig:performance}(b), we report the number of different solutions explored by our three approaches. Violet circles (green stars) [orange diamonds] refer to QTSA (Q-INSP) [SAMPLER]. Each line in the plot is the average, and the shadows represent the $95$\% confidence interval. Notice that, for each NTSP instance, we have chosen SAMPLER to evaluate a number of solutions close to the average number of points evaluated by QTSA
. 

As expected (see App.~\ref{sec:app_conf}), the number of solutions explored by Q-INSP is on average constant (approximately $2 \cdot 10^3$). On the other hand, QTSA explores more solutions for larger NTSP instances, where noise has a larger impact (see Sec.~\ref{sec:noise_mit_eval}). For all considered instances, QTSA and SAMPLER explore $\approx 100$ times more states than Q-INSP. When $n \leq 35$, the total number of possible solutions $2^n$ is sufficiently small to guarantee that QTSA and SAMPLER outperform Q-INSP. However, when $n \geq 40$ the classical optimizer of Q-INSP (see Sec.~\ref{sec:methodology}) is more effective at finding solutions. This approach achieves better payoff ratios $\rho$ than QTSA and, notably, yields $\rho$ values comparable to those obtained by SAMPLER, which evaluates a much larger (more than $100\times$) portion of the solution space.


%

In fact, since the output of the quantum circuit is on average a sparse state, Q-ISNP can approximate it well with the $10^4$ shots available at each iteration. As a consequence, the estimation of $C(\boldsymbol{\theta})$ in Eq.~\eqref{eq:final_obj} is sufficiently accurate for the classical solver to converge towards valid solutions (see Fig.~\ref{fig:convergence} below). 
%
On the other hand, QTSA operates in a 
regime where higher noise levels affect 
the estimation of the objective function in Eq.~\eqref{eq:final_obj}, undermining the parameters selection process. This favors the exploration that benefit the $n\leq 35$ instances but hinders the convergence that would be required for $n\geq 40$.
%

To further investigate the impact of hardware noise, we benchmark QTSA and Q-INSP with other hyperparameters configurations. 
Focusing on the NTSP instance of $20$ transactions (see Table~\ref{tab:data_stats}), we consider $5$ different scenarios: noiseless optimization (Q-INSP), and noisy optimization on both Eagle and Heron QPUs (QTSA) with and without iHAMMER. 
Five runs are conducted with Q-INSP and QTSA on Eagle QPUs, and three optimization runs on Heron QPUs (due to limited device access). We perform $100$ iterations per run, taking $2000$ measurement shots each and setting $\sigma = 0.14361$ in Eq.~\eqref{eq:mapping} to favor very sparse states (see App.~\ref{sec:app_conv_config}).
Additionally, we configure the BO acquisition function (see Sec.~\ref{sec:methodology}) to use the \emph{expected improvement} criterion (see App.~\ref{sec:app_bayesian_opt}). Compared to the experiments presented in Fig.~\ref{fig:performance}, this new parameters' configuration incentivizes faster convergence towards local minima.

The five columns in Fig.~\ref{fig:convergence}, organized into an upper and a lower plots, correspond to the scenarios introduced above (Q-INSP/QTSA, with/without iHAMMER). 
The violet lines display the average payoff $ C(\boldsymbol{\theta}) $ in Eq.~\eqref{eq:final_obj} at each iteration. The green dashed lines are $\max_\X C(\mathbf{x}, \mathbf{y}_\mathbf{x}) $, i.e., the highest payoff [see Eq.~\eqref{eq:activ_obj_fun}] found over all $\X$ explored up to that iteration. 
The orange lines represent the cumulative probability $ \sum_\X \tilde{p}_\X $ of the bitstrings $\X$ 
such that $ C(\X, \Y_\X) > 0 $. As explained in Sec.~\ref{sec:methodology}, this condition ensures that all constraints in Eqs.~\eqref{eq:nnb}-\eqref{eq:after_link} are satisfied by these candidate solutions. 

In Fig.~\ref{fig:convergence}(a) we present the results of the $5$ Q-INSP runs. Here (as in all plots), we highlight the run that most closely aligns with the expected convergence behavior of BO. This behavior is characterized by overall high (above zero) payoffs (violet) and a close-to-one cumulative probability function (orange). In other words, valid NTSP solutions must be reliably sampled from the output of the quantum circuit $\ket{\psi(\boldsymbol{\theta})}$, except for few exploratory iterations that are employed by BO to escape local minima (see App.~\ref{sec:app_bayesian_opt}). Convergence is observed in $4$ out of $5$ Q-INSP runs.


Fig.~\ref{fig:convergence}(b) includes the results obtained on Eagle QPUs (\texttt{ibm\_brisbane}) without iHAMMER. When operating in the noisy regime, erroneous $\X$ are sampled and the estimation of $C(\boldsymbol{\theta})$ in Eq.~\eqref{eq:final_obj} is compromised. 
As a result, in none of the $5$ QTSA runs performed on Eagle QPUs we observed a payoff trend line converging towards positive (i.e., valid) $C(\boldsymbol{\theta})$ values. Nonetheless, the cumulative probability of the highlighted run in Fig.~\ref{fig:convergence}(b) demonstrates how, even without a clear payoff convergence, BO selects parameters that consistently produce valid solutions with high probability. This is also highlighted by the best payoff (dashed green lines) found, which is always approaching the optimal (see Table \ref{tab:data_stats}).

In Fig.~\ref{fig:convergence}(c) we show the results obtained by QTSA, on Eagle and with iHAMMER. 
In two out of five runs, the payoff trend consistently surpass the $0$ threshold, resembling the desired convergence pattern observed in panel (a) by Q-INSP. However, even for the best run, the few 
erroneous bitstrings $\X$ that iHAMMER did not filter out lead to a less stable convergence compared to Q-INSP. The same observations hold true for the cumulative probability, which is correlated with the payoff. Notice how, compared to panel (b), the orange lines peak at values that are much closer to one. This clearly showcases the effectiveness of iHAMMER and its role in achieving convergence.

Finally, Figs.~\ref{fig:convergence}(d-e) show results obtained by running QTSA on Heron QPU (\texttt{ibm\_fez}). Due to limited access to the QPU, we were able to run only three QTSA instances with and without iHAMMER. Despite the improved noise resilience demonstrated in prior tests (see Sec.~\ref{sec:noise_mit_eval}), Heron devices did not showcase convergence.

\subsection{Discussion}
\label{sec:discussion}

The main goal of the experiments presented in Secs.~\ref{sec:noise_mit_eval} and \ref{sec:vqa_performance} is benchmarking the variational routine described in Sec.~\ref{sec:methodology} with state-of-the-art superconducting qubits quantum devices. 
As discussed in Sec.~\ref{sec:vqa_performance}, the noiseless VQA finds valid solutions to the NTSP (defined in Sec.~\ref{sec:model}) despite exploring only an exponentially small portion of the solution space. Indeed, Q-INSP works well for all the considered instances. In contrast, the QTSA solver appears to be significantly affected by hardware noise, preventing it from solving instances with more than $35$ transactions.

Despite the benefits provided by iHAMMER (see Sec.~\ref{sec:noise_mit_eval}), the results in Fig.~\ref{fig:convergence} suggest that more effective mitigation techniques and/or more noise-resilient QPUs are required to correctly estimate $C(\boldsymbol{\theta})$ in Eq.~\eqref{eq:final_obj} and enable convergence of the approach. The best known error mitigation techniques \cite{review_error_mitigation,chan2024measurement, rahman2022self_error_mitigation, urbanek2021_error_mitigation} 
correct the estimated values of physical observable. Instead, our optimization pipeline in Sec.~\ref{sec:methodology} is based on the probabilities $\tilde{p}_{\X}$. 

Correcting the probabilities $\tilde{p}_{\X}$ is a challenging task and a bottleneck of our VQA. To avoid this problem, a possibility is to express the objective function Eq.~\eqref{eq:final_obj} as a diagonal Ising Hamiltonian. However, while this can be achieved via, e.g., \cite{mohseni2022ising,lucas2014ising}, it would generally require an exponential number of terms, making this task unfeasible in practice.
Finding an efficient Hamiltonian encoding for the NTSP remains an intriguing open problem. Solving this would make it possible to substitute the ansatz in Sec.~\ref{sec:methodology} with the QAOA's ansatz in \cite{zhou2020quantum,lucas2014ising}.
At the expense of larger circuits' depths, this would provide performance guarantees (in terms of number of iterations) \cite{zhou2020quantum} that our protocol currently does not have. 

Hardware limitations are the main obstacle to the scalability of our approach. The first challenge encountered when working with NISQ devices is the limited number of available qubits. 
The size of the batches processed by T2S during run 4 of the night-time cycle is much larger
than the instances solved in this work using real quantum hardware (see Fig.~\ref{fig:performance}). Despite larger QPUs exist \cite{IBM2024roadmap}, based on Fig.~\ref{fig:fidelity} we expect that noise will be a limiting factor for these and the next generations of devices. This is taken into account by the IBM roadmap \cite{roadmap-ibm}, whose current focus is on noise reduction.


The results presented in Sec.~\ref{sec:vqa_performance} support the study in Ref.~\cite{scriva2024challenges}, where it is shown that sample and hardware noises make some VQAs perform similarly to a random sampler. Indeed, in Fig.~\ref{fig:performance} QTSA and SAMPLER yield results that are statistically compatible for NTSP instances with $\leq35$ transactions, and Fig.~\ref{fig:convergence} [except for panel (c)] demonstrates that BO does not converge when hardware noise is present. However, the two runs in Fig.~\ref{fig:convergence}(c) that show convergent behavior indicate how iHAMMER can provide (for small problem instances) the estimation accuracy that is required for BO to work as expected. This gives hope for future devices and error mitigation techniques.

\section{Conclusion and Outlook}
\label{sec:conclusion}

We have introduced a NISQ-compatible VQA for the NTSP modeled in Sec.~\ref{sec:model}. The quantum solver, whose architecture has been detailed in Sec.~\ref{sec:methodology}, has been evaluated over batches of transactions generated from real-world T2S data. Although tailored to the NTSP, the proposed methodology is applicable to a wide range of optimization problems that can be modeled as MIP.

NTSP instances with up to $35$ ($40$) transactions were solved by QTSA on gate-based superconducting quantum hardware (Q-INSP on classical machines). Furthermore, with iHAMMER we demonstrated convergence patterns for instances with $20$ transactions on real quantum hardware, thereby surpassing the size of NTSP instances successfully addressed by previous works \cite{braine2021quantum, huber2024exponential}. Our results demonstrate that, despite the challenges from noisy hardware, small-size NTSP instances can be solved on superconducting quantum devices.


As a first outlook, it would be interesting to extend the benchmark discussed in Sec.~\ref{sec:vqa_performance} to different hyperparameters (numbers of iterations, shots, BO settings, etc.; see App.~\ref{sec:app_conf}). This would enhance the performance of our approach and facilitate QTSA' convergence towards the desired solution. Second, it is important to develop other quantum algorithms for solving the NTSP problem. This includes more advanced VQA approaches whose ansatz could be based on the adiabatic theorem \cite{zhou2020quantum}, or exploit symmetries \cite{gunderman2024minimal}, as well as the integration of provably advantageous quantum methods \cite{kerzner2024square} in T2S' subroutines.



Advancing quantum algorithms to handle larger NTSP instances requires progress on both the hardware and algorithmic sides. This work provides a foundation for future studies that aim to optimize the use of NISQ devices for large-scale NTSP resolution.


\section{Acknowledgments}

This work was developed as part of a collaboration between the Bank of Italy, IBM, Intesa Sanpaolo, and academia. The authors recognize the invaluable support provided by all these entities throughout the study. In particular, we express our gratitude to Leonardo Di Paoloantonio and Giovanni D’Arrigo for their invaluable contributions in providing the data used in this work. 
We also thank
the IBM consulting team members Antonello Aita, Tommaso Fioravanti, Giulia Franco, Marco Masoero, and Mario Filadelfo Onorato.
LD acknowledges the EPSRC quantum career development grant EP/W028301/1 and the EPSRC Standard Research grant EP/Z534250/1.

\printbibliography

\clearpage
\onecolumngrid
\appendix

\section{Greedy Heuristic for Collateral Computation}
\label{sec:app_greedy}

We outlined in Algorithm~\ref{algo:greedy_coll} the classical heuristic developed for the computation of the $\Y$ decision variables of the NTSP. The main objective of this greedy algorithm is to estimate the maximum lots of securities that can be pledged through each $\spl \in \SPL$ while settling transactions according to a given $\X$. The algorithm relies on the Non-Shared Collateral assumption discussed in Table~\ref{tab:assumptions}, and achieves a time-complexity of $\mathcal{O}(T+BL)$, where $B$ is the number of cash balances, $L$ represents the number of CMB-security position links, and $T$ is the total number of transactions (we remark that the pseudo-code presented in Algorithm~\ref{algo:greedy_coll} is not optimized, having prioritized clarity over efficiency).

\begin{algorithm}
\caption{Collateral Computation Heuristic}
\label{algo:greedy_coll}
\KwIn{Settlement proposal $\X$, NTSP input data}
\KwOut{Pledgeable lots $\Y$}
$\Y \gets \mathbf{0}$ \tcp{Zero array with length equal to $|\SPL|$}

\For{each cash balance $\cbal$}{
    $val_\cbal \gets$ final balance in $\cbal$ after settling $\X$ \;
    \If{$val_\cbal > 0$}{
        \tcp{No additional credit required}
        \textbf{continue} \; 
    }
    \If{$-val_\cbal > \aLi_\cmb$}{
        \tcp{Credit required exceeds limit}
        \textbf{continue} \; 
    }
    $\Y_{tmp} \gets \mathbf{0}$ \tcp{Zero array with length equal to $\Y$}
    $credit \gets 0$ \;
    $\aLi_{tmp} \gets \aLi_\cmb$ \;

    \For{each $\spl \in \SPL$}{
        $val_\spos \gets$ final balance in $\spos$ after settling $\X$ \;
        $lot_{max} \gets$ max lot of securities pledgeable without exceeding $\aLi_{tmp}$ and $\qLi_\spl$ limits \;
        $lot_{value} \gets$ value of $lot_{max}$ in cash \;

        \If{$\qMi_\spl \leq lot_{max}$}{
            $\Y_{tmp}[\spl] \gets lot_{max}$ \;
            $credit \gets credit + lot_{value}$ \;
            $\aLi_{tmp} \gets \aLi_{tmp} - lot_{value}$ \tcp{Reduce limit}
        }
    }
    
    \If{$credit \text{ is enough, i.e., } val_\cbal + credit \geq 0$}{
        $\Y \gets \Y + \Y_{tmp}$ \tcp{Update collateral array}
    }
}
\Return $\Y$ \;
\end{algorithm}

\section{Iterative HAMMER}
\label{sec:app_ihammer}

As discussed in \cite{tannu2022hammer}, the outcomes of NISQ quantum circuits exhibit a structured behavior in Hamming space, where erroneous bitstrings are often close to the correct solution.
Building upon this evidence, the Hamming Reconstruction (HAMMER) mitigation technique assigns scores to bitstrings based on their frequency and their Hamming neighborhood's structure. This score is then utilized to enhance the probability of outcomes having rich neighborhoods and high sampling counts (likely to be correct), while reducing the likelihood of isolated, potentially erroneous bitstrings. In doing so, the algorithm does not take into account the circuit structure and the noise level affecting the quantum device, penalizing performance.  
For this reason, we extend the algorithm by introducing an iterative procedure that applies the HAMMER routine multiple times, depending on a circuit-dependent error coefficient.

Similar to the self-mitigation approach proposed in \cite{urbanek2021mitigating}, each iteration of the VQA involves two runs: a main run executing the ansatz 
with parameters set to $\boldsymbol{\theta}$, and a calibration run with the same circuit but parameters set to zero. By comparing the expected output from the calibration run (i.e. $\ket{0}$) with the actual measured result, we can determine the effective sampling rate $p_0^{\text{cal}}$ for the bitstring $0$. The error coefficient $\gamma$ is then defined as $\gamma = 1 - p_0^{\text{cal}}$, 
$\gamma = 1 - \bra{\psi_{cal}}\ket{0}\bra{0}\ket{\psi_{cal}}$.

This information is exploited to define the convergence criteria of the iterative procedure. 
Specifically, let $S$ be the number of samples obtained from the evaluation run, and $\tilde{p}_k=s_k/S$ the sampling ratio of bitstring $k$ sampled $s_k$ times, we compute the total number of samples absorbed after the $i$-th application of HAMMER as: 
\begin{equation}
    A_i = \sum_{\{k|\tilde{p}_k\neq0\}} \text{max}\{s_k \cdot (\tilde{p}_k - \tilde{p}_{k, i}) / \tilde{p}_k, 0\}, 
\end{equation}
where $\tilde{p}_{k, i}$ is the probability associated with the bitstring $k$ after applying HAMMER $i$ times.
The algorithm iterates until $ A_i \geq \frac{9}{10}S \gamma$, that is until the majority of the expected noisy samples have been removed. 
This redistribution process shifts probability weight from low-probability, noisy states to higher-probability states, effectively boosting the likelihood of correct outcomes beyond their initial levels.

\section{Bayesian Optimization}
\label{sec:app_bayesian_opt}

To implement the classical optimizer used throughout the experiments in Sec.~\ref{sec:vqa_performance}, we employ the \texttt{BayesianOptimization} Python library \cite{bayesianOptimizer}. Bayesian Optimization (BO) \cite{BO} is a sequential optimization strategy for black-box functions, particularly effective for non-convex functions that are expensive to evaluate and potentially noisy. This approach consists of two main components: a probabilistic model (also known as surrogate model), typically a Gaussian Process (GP), and an acquisition function.

The role of the probabilistic model is to approximate the objective function. Let $D = {(x_j, y_j)}_{j=1}^i$ be a set of observations, where $x_j$ are the input locations (corresponding to the model parameters under consideration), and $y_j$ are the observed outputs (i.e., our evaluation function).
A GP defines a distribution over possible functions that fit the data. The prediction at each possible location $x'$ is modeled as a mean $\mu(x')$ and a variance $\sigma^2(x')$, to represent the model's uncertainty. When the set of observations is extended, the model is updated accordingly.

The acquisition function guides the search process, based on the current probabilistic model. The acquisition function selects new points to evaluate by performing a trade-off between \textit{exploration}, corresponding to selecting points where there is high uncertainty for the objective function, and \textit{exploitation}, corresponding to select points for which the objective function is expected to be high.
Common acquisition functions are \textit{expected improvement}, \textit{probability of improvement}, and \textit{entropy search}. 

An advantage of using BO is its sample efficiency, requiring fewer function evaluations compared to many global optimization methods. This is highly beneficial in our case since estimating the target function on a quantum device is intrinsically costly. Furthermore, BO naturally incorporates noise in observations and does not require gradient information, making it suitable for the evaluation of noisy queries to non-differentiable functions.

However, BO faces several challenges. First, standard GP regression scales as $\mathcal{O}(i^3)$, where $i$ is the number of observations, limit the applicability of large scale evaluation. Second, the curse of dimensionality hinders BO from performing well in high dimensional spaces, due to the complexity of defining as a suitable class of surrogate models. 

Recent developments in BO research have focused on addressing this limitation (for example, by reducing the search space through Sparse Axis-Aligned Subspaces \cite{SAASBO}) and expanding its capabilities; this extension has not been explored yet and will be the subject of future studies.

\section{Experiments Configurations}
\label{sec:app_conf}

In this section, we report the different solvers' hyperparameters configurations used during the experiments discussed in Sec.~\ref{sec:experiment}. 

\subsection{Objective Function Hyperparameters}
\label{sec:app_obj_fun_conf}

The objective function presented in Eq.~\eqref{eq:activ_obj_fun} is influenced by the hyperparameter $\lambda$, which is part of the original NTSP objective function $\ofun(\X)$ (refer to Eq.~\eqref{eq:objf}), as well as by the non-linear activation functions \( h_v: \mathbb{R} \to \mathbb{R} \) that we introduce to include the NTSP constraints in Eq.~\eqref{eq:activ_obj_fun}.

\begin{figure}[h!t]
    \centering
    \includegraphics[width=0.49\linewidth]{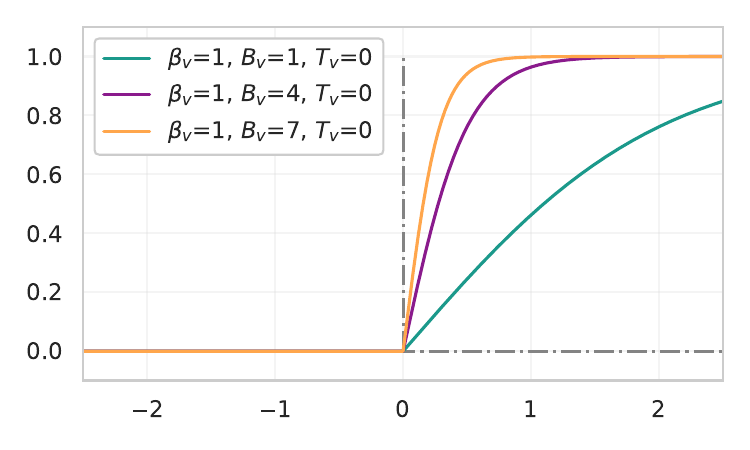}
    \includegraphics[width=0.49\linewidth]{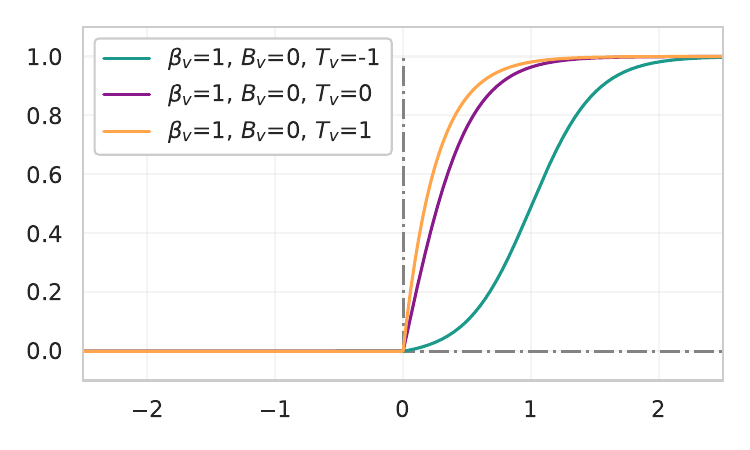}
    \caption{Different activation function $h_v$ (Eq.~\eqref{eq:hv}) generated by varying the parameters $B_v$ (on the left) and $T_v$ (on the right). Here, $\beta_v$ is fixed to $1$.}
    \label{fig:hv}
\end{figure}

We keep these hyperparameters fixed across all experiments. Specifically, $\lambda$ is set to $0.5$ to equally weight the volume and amount terms in Eq.~\eqref{eq:objf}. Each $h_v$ is selected from a family of functions defined by Eq.~\eqref{eq:hv}.
Given $\beta_v$, $T_v$, and $B_v$ for constraint $v$, and
\begin{equation}
    b_v(x) = \left(\frac{1}{ 1 + \exp\left( -B_v \cdot (x) \right) }\right), 
\end{equation}
all non linear activation function are defined as:
\begin{equation}
    \label{eq:hv}
    h_v(x) = 
        \begin{cases}
        0 & \text{if} \quad x \leq 0        \\
        \beta_v \frac{b_v(x) - b_v(T_v)}{1 - b_v(T_v)} & \text{otherwise.}
    \end{cases}
\end{equation}
We plot some of these activation functions in Fig~\ref{fig:hv}, changing the parameters $B_v$ and $T_v$.

For the experiment discussed in Sec.~\ref{sec:experiment}, we employed the hyper-parameterization described in Table \ref{tab:hv}.

\begin{table}[ht]
    \centering
    \begin{tabular}{cccc}
        \toprule
        \textbf{Constraint} & $\beta_v$ & $B_v$ & $T_v$ \\
        \midrule
        Eq.~\eqref{eq:nnb1} & $100$ & $0.025$ & $0$ \\ 
        Eq.~\eqref{eq:nnb2} & $10$ & $0.025$ & $0$ \\ 
        Eq.~\eqref{eq:after_link} & $5$ & $10$ & $-0.5$  \\
        \bottomrule
    \end{tabular}
    \caption{Hyperparameters chosen for the definition of the activation functions of the constraints in Eqs.~\eqref{eq:nnb1}\eqref{eq:nnb2}\eqref{eq:after_link}.  Note that the constraints in Eq.~\eqref{eq:cash_sec_coll} are not included into the objective function Eq.~\eqref{eq:final_obj}, as they are enforced by Algorithm~\ref{algo:greedy_coll} during the computation of $\Y_\X$.}
    \label{tab:hv}
\end{table}

\subsection{VQA Performance Analysis Experiment: Sec.~\ref{sec:vqa_performance}}
\label{sec:app_perf_config}

We list in Table \ref{tab:sigmas} the standard deviation values $\sigma$ chosen for the parameter mapping function $\text{G}(\theta_i)$ in Eq.~\eqref{eq:mapping}, and used in the experiment shown in Fig.~\ref{fig:performance}.
These values have been chosen by solving 
\begin{equation}
\label{eq:sigmas}
    P_{n, \sigma}(x \leq \lceil \log_2\delta \rceil) = 1-\frac{1}{S},
\end{equation}
for $\sigma \geq 0$, a given $n$ (i.e. number of qubits of the quantum circuit), $S$ (i.e. number of measurement shots at each iteration) and $\delta$ (i.e. expected number of bitstring measurable with non-negligible probability from the circuit output $\ket{\psi(\boldsymbol{\theta})}$). The function $P_{n, \sigma}(x \leq k)$ is defined as the cumulative distribution function of a binomial random variable $x$:
\begin{equation}
\label{eq:bin_sigmas}
P_{n, \sigma}(x \leq k) = \sum_{j=0}^k \binom{n}{j} \left(\frac{1 - e^{-\sigma/2}}{2} \right)^j \left( \frac{1 + e^{-\sigma/2}}{2} \right)^{n-j}
\end{equation}
Eq.~\ref{eq:sigmas} is ansatz-dependent, thus tailored for the parametrized quantum circuit described in Sec.~\ref{sec:methodology} and depicted in Fig.~\ref{fig:algo}.
\begin{table}[h!t]
    \centering
    \begin{tabular}{lccccc}
    \toprule
    & \textbf{gs\_20} &\textbf{gs\_25} &\textbf{gs\_30} & \textbf{gs\_35} & \textbf{gs\_40}  \\
    \midrule
    \textbf{Std. Dev. $\sigma$} & 0.35587 &   0.26877 &   0.21609 & 0.18074 & 0.15536 \\
    \bottomrule
    \end{tabular}
    \caption{Standard deviations values $\sigma$ obtained through Eq.~\eqref{eq:sigmas} for each NTSP size considered for the experiments in Fig.~\ref{fig:performance}. Here, we set $\delta=128$, and $S=10^4$.}
    \label{tab:sigmas}
\end{table}

In Table~\ref{tab:perf_exp} we list the hyperparameter configurations of Q-INSP and QTSA solvers used for the experiment shown in Fig.~\ref{fig:performance}. These solvers employed BO as parameters optimizer, with \emph{upper confidence bound} (ucb) as the acquisition function \cite{bayesianOptimizer}. 
\renewcommand{\arraystretch}{1.8}
\begin{table}[h!]
    \centering
    \begin{tabular}{l|c|c}
        \toprule
         & \textbf{Q-INSP} & \textbf{QTSA} \\
        \midrule
        \textbf{Backend} & - & [\texttt{ibm\_brisbane}, \texttt{ibm\_sherbrooke}] \\
        \textbf{Ansatz} & $R_y$ layer & $R_y$+C-NOTs layers \\
        \textbf{Shots} & $10^4$ & $10^4$ \\
        \textbf{Optimizer} & BO - ucb ($\kappa$: 2.576) & BO - ucb ($\kappa$: 2.576) \\
        \textbf{Iterations} & 300 & 300 \\
        \textbf{DD} & - & True \\
        \textbf{DD Sequence} & - & XpXm \\
        \textbf{iHAMMER} & - & True \\
        \bottomrule
        \end{tabular}
    \caption{Solvers' configurations employed in the experiments described in Fig.~\ref{fig:performance}.}
    \label{tab:perf_exp}
\end{table}

In Fig.~\ref{fig:performance} we discuss the results obtained by running the SAMPLER solver. Each run of this algorithm involves evaluating a fixed number of points that are sampled from the solution space. In Table~\ref{tab:samples} we write the number of samples considered by SAMPLER while tackling instances of different sizes: these have been chosen to match the average number of (different) solutions evaluated by QTSA.
\begin{table}[h!t]
    \centering
    \begin{tabular}{lccccc}
    \toprule
    & \textbf{gs\_20} &\textbf{gs\_25} &\textbf{gs\_30} & \textbf{gs\_35} & \textbf{gs\_40}  \\
    \midrule
    \textbf{Samples} & 282000 & 453000 & 676500 & 925500 & 1080000 \\
    \bottomrule
    \end{tabular}
    \caption{Number of random samples evaluated by SAMPLER, distinguishing between instances of different size.}
    \label{tab:samples}
\end{table}

\subsection{VQA Convergence Experiment: Sec.~\ref{sec:noise_mit_eval}}
\label{sec:app_conv_config}

In Table~\ref{tab:convergence_exp} we list the hyperparameter configurations used in the experiment shown in Fig.~\ref{fig:convergence}. In contrast with previous experiments, we chose here \emph{expected improvement} (ei) as the acquisition function for BO \cite{bayesianOptimizer}. The value of $\sigma$ was obtained by solving Eq.~\eqref{eq:sigmas} with $S=2000$, $n=20$, and, notably, $\delta=16$.   

\begin{table}[h!t]    
    \centering
    \begin{tabular}{l|ccccc}
        \toprule
         & \multicolumn{1}{c}{\textbf{Noiseless}} & \multicolumn{2}{|c|}{\textbf{Eagle QPUs}} & \multicolumn{2}{c}{\textbf{Heron QPUs}} \\
        \midrule
        \textbf{iHAMMER}    & - & \multicolumn{1}{|c}{$~~$False$~$} & \multicolumn{1}{|c|}{True} & \multicolumn{1}{c}{$~~$False$~$} & \multicolumn{1}{|c}{True} \\
        \midrule
        \textbf{Solver}     & Q-INSP    &   \multicolumn{2}{|c|}{QTSA} & \multicolumn{2}{c}{QTSA} \\
        \textbf{Instance}   & gs\_20    &   \multicolumn{2}{|c|}{gs\_20} & \multicolumn{2}{c}{gs\_20} \\
        \textbf{Backend}    & -         &   \multicolumn{2}{|c|}{\texttt{ibm\_brisbane}} & \multicolumn{2}{c}{\texttt{ibm\_fez}} \\
        \textbf{Ansatz}     & $R_y$ layer & \multicolumn{2}{|c|}{$R_y$+C-NOTs layers} & \multicolumn{2}{c}{$R_y$+C-NOTs layers}  \\
        \textbf{Shots}      & 2000      &   \multicolumn{2}{|c|}{2000} & \multicolumn{2}{c}{2000} \\
        \textbf{Optimizer}  & BO - ei ($\xi$: 0.75) & \multicolumn{2}{|c|}{BO - ei ($\xi$: 0.75)} & \multicolumn{2}{c}{BO - ei ($\xi$: 0.75)} \\
        \textbf{Iterations} & 100       & \multicolumn{2}{|c|}{100} & \multicolumn{2}{c}{100} \\
        \textbf{Std. Dev. $\sigma$}  & 0.14361   & \multicolumn{2}{|c|}{0.14361} & \multicolumn{2}{c}{0.14361} \\
        \textbf{DD}         & -         & \multicolumn{2}{|c|}{True} & \multicolumn{2}{c}{True} \\
        \textbf{DD Sequence} & -        & \multicolumn{2}{|c|}{XpXm} & \multicolumn{2}{c}{XpXm} \\
        \bottomrule
    \end{tabular}
    \caption{Solvers' configurations employed in the experiments described in Fig.~\ref{fig:convergence}.}
    \label{tab:convergence_exp}
\end{table}

\end{document}